%% file: main.tex
\documentclass[prl,twocolumn,english,superscriptaddress,floatfix,longbibliography,aps,amsmath,amssymb]{revtex4-2}

\input{preamble}

\usepackage[utf8]{inputenc}
\usepackage{xcolor}
\usepackage{enumitem}

\begin{document}
\title{Generalized Time-Coarse Graining via an Operator Cumulant Expansion}
\author{Leon Bello}
\email[Correspondence email address: ]{leon.bello@weizmann.ac.il}
\affiliation{Department of Electrical and Computer Engineering, Princeton University}
\author{Tal Rubin}
\affiliation{Department of Astrophysical Sciences, Princeton University}
\author{Wentao Fan}
\affiliation{Department of Electrical and Computer Engineering, Princeton University}
\author{Nathaniel Fisch}
\affiliation{Department of Astrophysical Sciences, Princeton University}
\author{Hakan E. Türeci}
\affiliation{Department of Electrical and Computer Engineering, Princeton University}

\date{\today} 

\begin{abstract}
We introduce a general framework for deriving effective dynamics from arbitrary time-dependent generators, based on a systematic operator cumulant expansion. Unlike traditional approaches, which typically assume periodic or adiabatic driving, our method applies to systems with general time dependencies and is compatible with any dynamics generated by a linear operator—Hamiltonian or not, quantum or classical, open or closed. This enables  modeling of systems exhibiting strong modulation, dissipation, or non-adiabatic effects. Our approach unifies Hamiltonian techniques such as Lie-transform Perturbation Theory (LPT) with averaging-based methods like Time-Coarse Graining (TCG), revealing their structural equivalence through the lens of generalized cumulants. It also clarifies how non-Hamiltonian terms naturally emerge from averaging procedures, even in closed systems. We illustrate the power and flexibility of the method by analyzing a damped, parametrically driven Kapitza pendulum, a system beyond the reach of standard tools, demonstrating how accurate effective equations can be derived across a wide range of regimes.
\end{abstract}


\maketitle
\label{sec:intro} 

Driven systems are fundamental to many areas of physics, from plasma physics to quantum optics, giving rise to rich phenomena such as chaos, parametric instabilities, pattern formation \cite{perego_pattern_2016}, and synchronization \cite{tarasov_mode-locking_2016, bello_persistent_2019}. Their complex and often counterintuitive dynamics underpins a plethora of theoretical and practical applications; classical plasma instabilities and pondermotive forces \cite{barth_ladder_2015, rubin_flowing_2024}; the generation of highly non-classical quantum states of light \cite{leghtas_confining_2015, lescanne_exponential_2020, grimm_stabilization_2020}; and emergent Floquet phases in periodically driven many-body systems \cite{wintersperger_realization_2020}. The interplay between external drives and nonlinear effects typically leads to intricate multi-scale dynamics that lack closed-form solutions, and are challenging to study even numerically~\cite{landau_mechanics_1976,arnold_mathematical_2009}. Developing methods to effectively model such systems remains a fundamental problem across multiple domains of physics and engineering. Despite the ubiquity of time-dependent systems, existing analytical methods are typically tailored to narrow regimes—such as periodic driving, high-frequency expansions, or adiabatic modulation—where the structure of the solution is well understood \cite{jackson_perspectives_1990}. While these tools are often effective and interpretable within their domains, extending them to more general time dependencies or dissipative settings quickly becomes cumbersome, opaque, or non-systematic. As a result, they offer limited guidance in analyzing complex, non-adiabatic, or transiently driven systems that arise in many modern applications.

\begin{figure}[h!]
    \centering
    \includegraphics[width=0.75\linewidth]{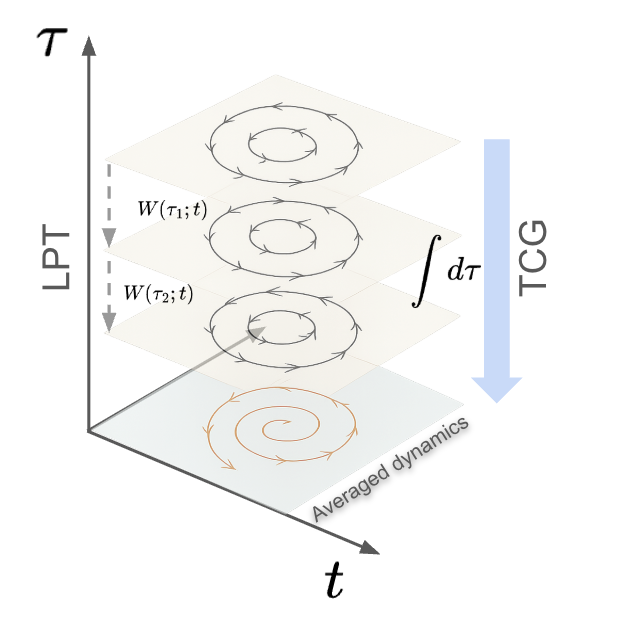}
    \caption{Illustration of Hamiltonian vs. averaging methods. The $t$ and $\tau$ axes denote slow and fast time-scales. Hamiltonian methods split phase space into slow and fast parts, each governed by its own Hamiltonian. Averaging methods smooth over the fast scale, yielding non-Hamiltonian dynamics.}
    \label{fig:tcg_vs_lpt}
\end{figure}

This work introduces a general framework for deriving an effective generator for the time-averaged dynamics in time-dependent systems, providing a unifying perspective on transformation-based and averaging-based approaches and extending them beyond the state-of-the-art. Our approach applies broadly to systems generated by time-dependent linear operators—Hamiltonian or not, quantum or classical, open or closed—without requiring periodicity, adiabaticity, or other simplifying assumptions. This flexibility allows us to tackle strongly non-adiabatic and non-periodic systems that lie beyond the reach of traditional methods; As an illustrative example, we solve a damped, parametrically modulated Kapitza pendulum\cite{kapitza__1951,stephenson_xx_1908}, and demonstrate results that go beyond the current state-of-the-art\cite{Blackburn_1992AmJPh, Butikov2001}. We formulate the method using an abstract but intuitive framework built on ideas from non-commutative probability and algebraic structures such as generalized cumulant expansions~\cite{kubo_generalized_1962,bonnier_signature_2020,lehner_cumulants_2011}, path signatures~\cite{friz_course_2020,chevyrev_signature_2018,chevyrev_signature_2018}, and free Lie algebras~\cite{reutenauer_free_1993,lehner_free_2002,speicher_free_2009}. This perspective provides a unified algebraic language that connects quantum and classical systems, and bridges disparate techniques under a common formalism. It not only clarifies deep structural similarities between these approaches, but also extends their applicability to a far broader class of dynamical systems.

\emph{General formulation.} -- Most physical systems of interest can be described by equations of motion which can generally be expressed in terms of a linear generator: 
\begin{equation}
    \frac{d}{dt}{z} = \mathcal{L}(t) z
\end{equation}
where $\mathcal{L}(t)$ is a (possibly time-dependent) linear operator. Note that while $\mathcal{L}(t)$ is a linear operator over the observable $z$, the resulting equations of motion may still be nonlinear functions of the phase-space variables themselves. In Hamiltonian systems, the generators are Lie derivatives defined by system-specific Lie brackets $\dot{z} = \mathcal{L}_H z = \{\{z, H\}\}$, where the Lie-bracket $\{\{A, B\}\}$ is any bilinear, antisymmetric operation satisfying the Jacobi identity. For example, Poisson brackets in classical mechanics \cite{goldstein_classical_2008}, Heisenberg commutators in quantum mechanics \cite{sakurai_advanced_1967}, and Moyal brackets in quantum quasi-probability distributions \cite{moyal_quantum_1949}. Open Hamiltonian systems extend this linear framework, incorporating dissipation through additional linear operators such as Vlasov-type double-brackets \cite{morrison_thoughts_2009}  or Lindblad-type dissipation operators \cite{lindblad_generators_1976}, thus preserving linearity despite non-conservative dynamics. 

Formally, the exact time evolution $z(t) = \mathcal{S}(t)z_0$ of such systems is given by the time-ordered exponential map,
\begin{equation}
    \mathcal{S}(t) = \exp_t \left( \int_0^t dt'\, \mathcal{L}(t') \right) 
\end{equation}
where $\exp_t$ denotes time-ordered exponential. Although easily formulated, the calculation of this propagator is generally intractable. 

A common strategy for analyzing such time-dependent systems is to separate fast and slow degrees of freedom \cite{maggia_higher-order_2020,bukov_universal_2015}. Two main paradigms exist: Hamiltonian methods like Lie-perturbation theory (LPT), which isolate slow motion via canonical transformations, and averaging-based methods like time-coarse graining (TCG), which smooth over fast scales. Figure~\ref{fig:tcg_vs_lpt} illustrates the difference between the two approaches: In LPT, a sequence of near-identity transformations separates the Hamiltonian into slow and fast components, producing an effective Hamiltonian that governs the coarse-grained dynamics~\cite{cary_lie_1981,dewar_renormalised_1976}. This technique is widely used in both classical and quantum contexts and has recently been unified within a Lie-algebraic framework~\cite{venkatraman_static_2022}. By contrast, TCG is a uniquely quantum approach that defines a time-averaged density operator $\overline{\rho}$ and derives its dynamics directly. The resulting generator typically includes non-Hamiltonian terms resembling Lindblad evolution—even in closed systems~\cite{bello_systematic_2025,fan_model_2024,gamel_time-averaged_2010}. While successful in quantum applications, the connection between TCG and transformation-based methods like LPT has remained unclear, and the structure of its non-Hamiltonian corrections lacks a unified interpretation.

\begin{figure}[!h]
    \centering
    \includegraphics[width=0.92\linewidth]{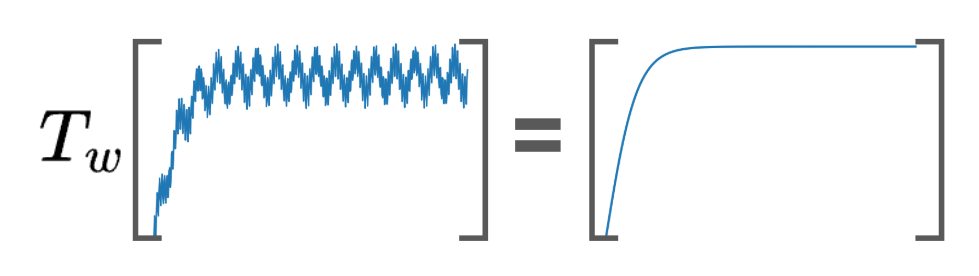}
    \caption{$T_w$ should act like a filtering operator, retaining only the dynamics of interest.}
    \label{fig:averaging_schematic}
\end{figure}

We propose a unifying perspective in which seemingly disparate methods are recast as linear, invertible \emph{filtering transformations} $T_w$, parametrized by a filter function $w(t)$. As illustrated in Figure~\ref{fig:averaging_schematic}, each transformation $T_w$ is an averaging-like operation that suppresses unwanted time-scales, yielding a simplified representation of the slow dynamics. For brevity, and to emphasize its connection to averaging, we write $T_w \square \equiv \overline{\square}$. 

For example, in \textit{Lie-transform perturbation theory}, $T_w$ corresponds to a canonical transformation generated by a Hamiltonian-like function $w(t)$, that isolates a slow effective Hamiltonian generator (known in the literature as the ``Kamiltonian'' or ``gyro-center Hamiltonian''). In contrast, \textit{time-coarse graining} (TCG) uses a convolution operator as $T_w$, producing non-canonical evolution equations that naturally include non-Hamiltonian terms, where $w(t)$ is a filter function that removes the dynamics outside a specific band of interest. These two cases, traditionally seen as distinct, emerge here as structurally unified under our filtering framework.

\emph{Cumulant-expansion derivation.} -- Since the filtering transformation is linear, the dynamics in the transformed frame is still governed by an effective generator (see the supplementary material for more details):
\begin{equation}
    \frac{d}{dt} \overline{z}(t) = \mathcal{U} \, \overline{z}(t),
\end{equation}
where \(\mathcal{U}\) depends implicitly on both the original generator $\mathcal{L}(t)$ and the filtering transformation $T_w$. With a suitable choice of \(T_w\), the effective generator \(\mathcal{U}\) becomes significantly simplified—often slowly varying, or even time-independent—making the dynamics in the filtered frame easier to analyze and solve. 

As in the original frame, we can define the formal propagator for the time-averaged dynamics. This time, instead of writing it as a time-ordered exponential, we will write the generator $\mathcal{U}(t)$ as the log-derivative of the propagator $\mathcal{M}(t)$,
\begin{equation}
    \label{eq:tcg_propagator}
    \frac{\rm d}{\rm{d}t} \mathcal{M}(t) = \mathcal{U}(t) \mathcal{M}(t).
\end{equation}
The averaged propagator is $\mathcal{M}(t) \equiv \overline{\mathcal{S}(t)}$, which plays the role of a time-ordered moment-generating function in the sense of Kubo et al.~\cite{kubo_generalized_1962}.
\begin{equation}
    \mathcal{M}_n(t) = \overline{\int_0^t dt_1 \int_0^{t_1} dt_2 \cdots \int_0^{t_{n-1}} dt_n \mathcal{L}(t_1) \cdots \mathcal{L}(t_n)} 
\end{equation}
This defines the effective generator $\mathcal{U}(t)$ as a \emph{cumulant-generating function}. Since operators do not commute, the definition depends on ordering; here we use the time-ordered exponential. In the Supplementary Material we also present an unordered version, which connects directly to the Magnus expansion and reduces to the standard cumulant-generating function in the commutative limit.

Our goal is to find an expression for the generator $\mathcal{U}$ of the time-averaged dynamics $\overline{z}(t)$ by solving the associated log-derivative equation. While this equation cannot generally be solved exactly, it can be inverted perturbatively by expanding both $\mathcal{U}(t)$ and $\mathcal{M}(t)$ in a formal power series, yielding explicit expressions for the ordered moments and cumulants.
\begin{equation}
\label{eq:tcg_generator}
	\mathcal{U}_n = \sum_{k=1}^n (-1)^{k+1} \sum_{(n_1,\ldots,n_k) \in \text{Comp}_k(n)} \dot{\mathcal{M}}_{n_1} \mathcal{M}_{n_2} \cdots \mathcal{M}_{n_k}
\end{equation} 
where the sum is over the set of $k$ partitions such that $\sum_{i=1}^k n_i = n$. 

Eq.~\ref{eq:tcg_generator} is the central result: for any filter $T_w$, the filtered variables $\overline{z}$ evolve under the cumulant-rate average of the generators. This formulation is broadly applicable—to classical or quantum systems, with diverse transformations—and provides a systematic, non-recursive way to compute corrections.

\emph{Generators with multiplicative time dependence}. -- In principle, the framework above allows $\mathcal{U}$ to be computed to arbitrary order, but direct calculations quickly become intractable. A powerful simplification arises when the generator has the form:
\begin{equation}
    \mathcal{L}(t) = \sum_{i \in \Omega} f_i(t)\,\mathcal{L}_i,
\end{equation}
where $\mathcal{L}_i$ are time-independent operators and all explicit time dependence is carried by scalar functions $f_i(t)$. Many physical systems naturally take this form, and it substantially reduces the algebraic complexity.

With this structure, the ordered moments reduce to
\begin{equation}
    M_w = \overline{\int_0^t dt_1 \int_0^{t_1} dt_2 \cdots \int_0^{t_{n-1}} dt_n f_{w_1}(t_1)\cdots f_{w_n}(t_n)},
\end{equation}
which coincide with the \emph{expected signatures} of stochastic process theory~\cite{bonnier_signature_2020}. These form a systematic coordinate system for ordered paths and the building blocks of our expansion. For concreteness, we use a convolution moving average as in time-coarse graining,
$\overline{\square}(t) = \int_{-\infty}^\infty w(\tau)\,\square(t-\tau)\,d\tau$
a convenient and physically transparent choice, though our results are independent of it.

In this setting, the corrections to the effective generator $\mathcal{U}(t)$ appear as products of the elementary generators  $L_w = \mathcal{L}_{w_1}\cdots \mathcal{L}_{w_n}$, which we call \emph{words},
\begin{equation}
    \mathcal{U}_n = \sum_{w \in \Omega^n} U_w L_w 
\end{equation}
Each word defines an ordered product of operators $w = (w_1,\dots,w_n)$, and we also define the concatenation of two  $wv = w_1\cdots w_n v_1\cdots v_m$. The coefficients $U_w$ are known as the signature cumulants \cite{bonnier_signature_2020},
\begin{equation}
    U_w =  \sum_{\pi \in \mathcal{P}(w)} \sum_{\underline{w} \in \pi(w)} (-1)^{|\underline{w}|+1} \dot{M}_{\underline{w}_1} \cdots M_{\underline{w}_{|\underline{w}|}}
\end{equation}
where the set $\mathcal{P}(w)$ is the set of all ordered partitions of the word \cite{bonnier_signature_2020,friz_unified_2021}. 

More explicit forms can only be given by choosing a specific time-dependence for $f_{i}(t)$. In previous work \cite{bello_systematic_2025}, we derived this for a simple harmonic time-dependence $f_i(t) = f_{\omega_i}(t) = e^{-i\omega_i t}$ in the context of quantum systems, deriving general closed-form formulas. We discuss this specific case in more detail in the supplementary material.

\emph{Dissipative and Hamiltonian contributions}. -- Hamiltonian systems, quantum or classical, are generated by Lie derivative. The abstract algebra, referencing only to the general algebraic properties of the Lie bracket, is called the Free Lie Algebra \cite{reutenauer_free_1993}. However, the averaged generator $\mathcal{U}$ is generally not a Lie-derivative, even when the original problem is fully Hamiltonian. 

In such cases, where the original problem is a closed Hamiltonian system, we can separate the average generator into a Lie contribution (generated by an effective Hamiltonian), and a non-Lie contribution. Given a word in the original free algebra, we can extract its Lie component using the Dynkin Idempotent \cite{patras_dynkin_2002, reutenauer_free_1993}, which projects the word from universal algebra to the associated free Lie algebra, and satisfies $\Theta_n^2[L_w] = \Theta_n[L_w]$. 
\begin{subequations}
\label{eq:hamiltonian_component}
 \begin{equation}
	\Theta_n[L_w] = \frac{1}{n} [\mathcal{L}_{w_1}, [\mathcal{L}_{w_2}, [\cdots [\mathcal{L}_{w_{n-1}}, \mathcal{L}_{w_n}]\cdots]]]    
\end{equation}
\begin{equation}
    \mathcal{H}_w = \frac{1}{n}\{H_{w_1}, \{H_{w_2}, \{\cdots\{H_{w_{n-1}}, H_{w_n}\} \cdots \} \} \}
\end{equation}
\end{subequations}
where Eq.~\ref{eq:hamiltonian_component}b is the associated Hamiltonian generator, arising directly from the Lie identity, $[\mathcal{L}_A, \mathcal{L}_B] = \mathcal{L}_{\{A, B\}}$.

In addition to the Hamiltonian terms, there will be a non-Hamiltonian remainder, arising from averaged product terms that break the simple Lie-derivative structure of the problem. These terms are fundamentally non-Hamiltonian and cannot be written as a simple Lie-derivative. They arise from the truncation of words, which breaks the invertibility of the averaging transformation, and consequently the Lie-algebraic structure of the generator~\cite{friz_unified_2021}.
They generalize the well-known Lindblad operators  \cite{lindblad_generators_1976,manzano_short_2020} and Vlasov double-brackets \cite{morrison_thoughts_2009}, and capture the effective loss of information and energy due to averaging (or alternatively, due to incomplete measurements~\cite{bello_systematic_2024, fan_model_2024}).


\emph{Example –- Damped Kapitza Pendulum.} The Kapitza pendulum is a vertically driven extension of the simple pendulum ~\cite{kapitza__1951, stephenson_xx_1908}, described by the Hamiltonian:
\begin{equation}
H(\theta,p_{\theta},t) = \frac{p_{\theta}^{2}}{2m l^{2}} + m l \left( g - \lambda l \nu^{2} \cos \nu t \right) \cos\theta,
\end{equation}
where $\theta$ is the angle from the inverted position, and $p_\theta$ is the conjugate momentum. The length of the pendulum is $l$ and its pivot is driven vertically at frequency $\nu$. 

The Kapitza pendulum exhibits rich dynamics including dynamical stabilization, parametric resonances, and chaos, with a wide range of experimental demonstrations \cite{bhadra_dynamics_2020,kulikov_kapitza_2022}. It is a hallmark problem for demonstrating dynamical stabilization and its properties are well-studied in the ideal, dissipation-free limit. However, analytical treatments that include damping are far more limited \cite{Blackburn_1992AmJPh,brockett1991dynamical}. Since our framework works for general linear operators, it naturally incorporates dissipation, making such analyses straightforward. In order to incorporate dissipation, we use the double-bracket dissipator formalism~\cite{brockett1991dynamical, Vallis_Carnevale_Young_1989},
\begin{equation}
    D(\theta,p_{\theta}) = -\frac{1}{2} Q^{-1} \omega_0 p_{\theta}^{2}, \quad \omega_0 = \sqrt{g/l},
\end{equation}
where $Q$ is the quality factor of the resonator.

\begin{figure}[h!]
    \centering
    \includegraphics[width=0.8\linewidth]{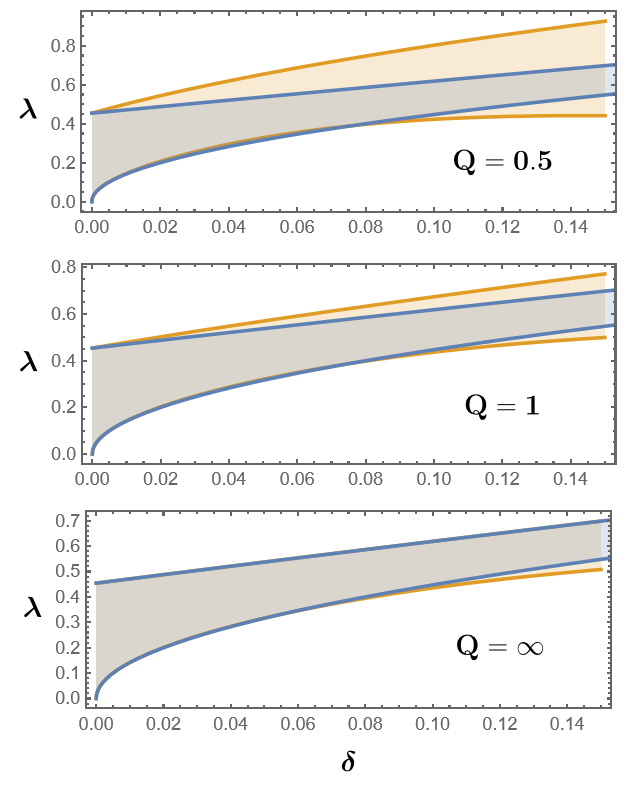}
    \caption{Shaded regions indicate stable inverted pendulum regimes. Blue lines show standard analytical boundaries; orange lines are computed from our seventh-order effective generator.}
    \label{fig:Kapitza_regime of dynamical stabilization}
\end{figure}

We can now apply our method to the damped Kapitza pendulum and determine the stability phases of the inverted position and derive accurate critical drive thresholds, providing results beyond the best available calculations in the literature. The first order of our method is equivalent to standard averaging of the generator, simply removing the oscillatory term in $H$, and yielding the following equations of motion,
\begin{equation}
    \partial_t \theta = \nu \tilde{p}_\theta, \quad \partial_t \tilde{p}_\theta = \nu(\Gamma^2 \theta - \beta \tilde{p}_\theta),
\end{equation}
with \(\tilde{p}_\theta = p_\theta/(ml^2\nu)\), \(\Gamma^2 = \omega_0^2/\nu^2\), and \(\beta = \Gamma Q^{-1}\). This predicts instability of the inverted position, as expected from standard treatments of the Kapitza pendulum. Stabilization appears only at higher orders: by symmetry, even order contributions vanish and the leading correction arises at third order. In the Supplementary Material we derive the effective generator up to seventh order, which allows us to calculate the stabilization threshold up to 7-th order in the perturbation parameters,
\begin{equation}
    \left(\lambda_0^{(7)}\right)^2 \approx 2\Gamma^2 \left(1 + \tfrac{7}{8}\Gamma^2 + \beta^2 - \tfrac{575}{32}\Gamma^4 - \tfrac{75}{8}\Gamma^2 \beta^2 \right),
\end{equation}
which refines the standard estimate $\lambda_0 \approx \sqrt{2\Gamma^2}$ \cite{Blackburn_1992AmJPh} and incorporates dissipation-induced corrections, which naturally arise using our method.

At higher drive strengths, the inverted fixed point destabilizes into a limit cycle at frequency $\nu/2$, with critical value $\lambda_c$. Using a self-consistent two-frame analysis and computing the fifth-order generator, we calculate an estimate for this parametric instability,
\begin{equation}
\lambda_c^{(5)} \approx 0.454 + 1.682\Gamma^2 - 0.413\Gamma^4  -2.62\beta^2 - 0.983\beta^4,
\end{equation}
which goes beyond that standard analytical estimate \cite{Blackburn_1992AmJPh,Butikov2001}. The full calculation is given in the supplementary material.
The stability boundaries at different orders are shown in Fig.~\ref{fig:Kapitza_regime of dynamical stabilization}, where the shaded region shows the stability phase diagram. Our method captures significant corrections to both lower and upper critical drive strengths, particularly at moderate detunings $\Gamma$ in the presence of dissipation, well beyond what is accessible via standard methods.

\emph{Example -- Length-modulated Kapitza Pendulum.} A key strength of our framework is its ability to capture non-adiabatic modulations, which are notoriously difficult using standard approaches. Floquet theory, for instance, requires strictly periodic driving and cannot account for slowly varying or transient changes, while high-frequency expansions rely on adiabatic assumptions that break down in such regimes. Our framework is robust, and is able to handle arbitrary time-dependencies, able to systematically produce the effective Hamiltonian and dissipator corrections order by order, enabling controlled approximations in regimes inaccessible to Floquet or high-frequency expansions. To illustrate this advantage, we consider a Kapitza pendulum whose length is slowly modulated, as described by the Hamiltonian:
\begin{equation}
\begin{split}
H(t)
=&
\frac{p_{\theta}^{2}}{2m l_{0}^{2}}
-
\frac{p_{\theta}^{2}}{m l_{0}^{2}} \Delta(t)\\
&
+
m g l_{0} \left( 1 + \Delta(t) \right)
\left(
1 + \lambda \cos( \nu t )
\right)
\cos\theta,
\end{split}
\end{equation}
where the modulation strength $\Delta(t) \equiv \alpha_{1} t + \alpha_{2} t^{2}$  is weak and changes slowly compared to the averaging timescale $\tau$ and the main modulation rate $\nu$. As in the earlier example, we also assume dissipation, governed by the same dissipator as in the previous example.

\begin{figure}[h!]
    \centering
    \includegraphics[width=0.8\linewidth]{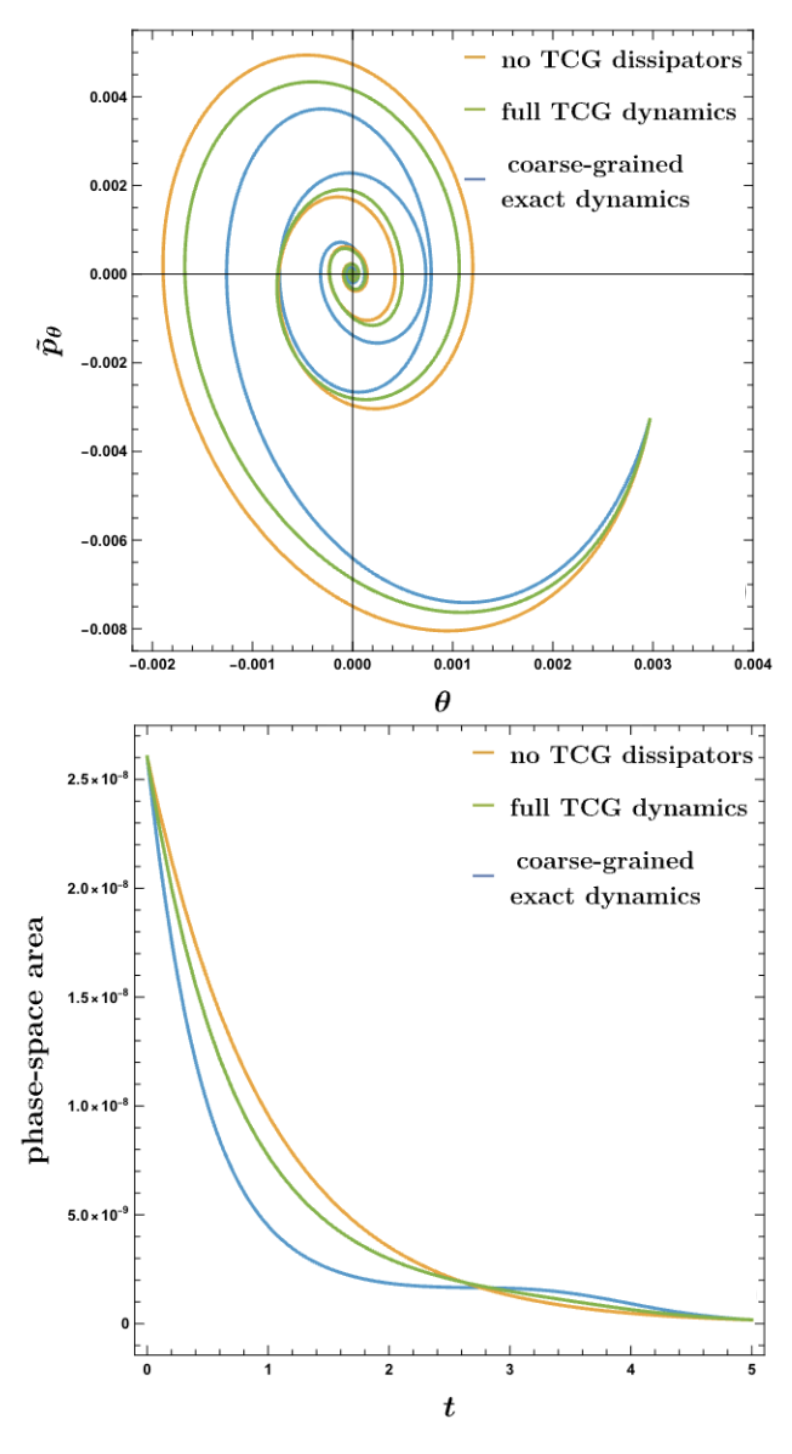}
    \caption{\textbf{(top)} Phase-space dynamics obtained from the exact (numeric) time-coarse grained dynamics versus the effective time-coarse grained dynamics obtained using our method. \textbf{(bottom)} The phase-space area obtained using the method. The time averaging causes the phase-space area to shrink, a manifestly non-Hamiltonian effect. Our numerics assume the following model parameters: $\Gamma^2 = 0.02, \lambda = 0.3, \nu = 20, \beta = 0.05, A = 0.2, T = 5/2\pi$, with a time-coarse graining scale of $\tau=0.4$.}
    \label{fig:phase-space_kapitza}
\end{figure}

Our method allows us to systematically derive corrections to the effective Hamiltonian and dissipators, order by order.  We compare the phase-space dynamics $(\theta, \tilde{p}_\theta) \equiv (\theta, \frac{p}{ml^2(t)})$ between two approaches. In the first, we calculate the effective generator perturbatively and use it to calculate the phase-space evolution. In the second, we solve the example numerically and then filter the dynamics. In Fig. \ref{fig:phase-space_kapitza}, we show how the phase-space trajectories of the two approaches match for different orders of the perturbation.

To first-order, the corrections are simply the time-averaged generators, or time-averaged Hamiltonian and dissipator,
\begin{equation}
\begin{split}
H^{(1)}_{\textrm{eff}}(t)
=
\overline{H}(t)
,\qquad
D^{(1)}_{\textrm{eff}}(t)
=
\overline{D}(t)
\end{split}
\end{equation}

Beyond first-order, the corrections cannot be fully captured by a simple effective Hamiltonian, and the contributions are have both Hamiltonian and non-Hamiltonian components. For example, in second-order, we see such non-trivial contributions arising from the non-adiabatic modulation of the pendulum length,
\begin{equation}
\begin{split}
&
H^{(2)}_{\textrm{eff}}(t)
=
\Gamma^2 \tau^{2} \nu^{2} \overline{\dot{\Delta}(t)}
\left[
\left( \sin\theta \right) p_{\theta}
-
\beta \nu m l_{0}^{2} \cos\theta
\right]\\
&
D^{(2)}_{\textrm{eff}}(t)
=
-
\frac{1}{2} \Gamma^2 \tau^{2} \nu^{2}
\overline{\dot{\Delta}(t)} \cos\theta
p_{\theta}^{2}
\end{split}
\end{equation}
these terms can be seen as fully non-adiabatic, since the vanish in the limit $\alpha_1, \alpha_2 \rightarrow 0$. Remarkably, these correction  persist even when the original problem is completely energy-conserving ($\beta \to 0$). Thus, non-adiabatic modulation combined with averaging generates effective dissipation absent in the original dynamics. 

At third-order we start seeing contributions from higher-derivatives of the modulation,
\begin{equation}
\begin{split}
&
H^{(3)}_{\textrm{eff}}(t)\\
=&
-
\Gamma^2 \alpha_{2} \nu^{2} \tau^{4}
\left(
\left( \cos\theta \right) \frac{p_{\theta}^{2}}{m l_{0}^{2}}
+
\beta \nu \left( \sin\theta \right) p_{\theta}
\right)\\
&
+
m l_{0}^{2} \Big[
\alpha_{2} \beta^{2} \Gamma^2 \nu^{4}\tau^{4} \cos\theta\\
&
-
\left(
\frac{\lambda^{2} \nu^{2}}{8}
+
\frac{\overline{\Delta(t)}\lambda^{2}\nu^{2}}{4}
-
\alpha_{2}\left( \lambda^{2} + \frac{\Gamma^4\nu^{4} \tau^{4}}{4} \right)
\right) \cos(2\theta)
\Big]\\
&
D^{(3)}_{\textrm{eff}}(t)
=
\left[
\frac{\alpha_{2} \Gamma^2 \nu^{2} \tau^{4}}{2}
\left(
\beta \nu \left( \cos\theta \right) p_{\theta}^{2}
+
\left( \sin\theta \right) \frac{2p_{\theta}^{3}}{m l_{0}^{2}}
\right)
\right]
\end{split}
\end{equation}
Note that although in this example the corrections separate neatly into Hamiltonian and dissipative parts, in general they  take more intricate forms beyond the simple double-bracket structure. 
Moreover, the corrections are generally dependent on the choice of filter, and in particular, the time-coarse graining scale $\tau$. However, some terms are independent of $\tau$ and survive even in the $\tau \rightarrow 0$ limit.

\label{sec:conclusions}
\emph{Conclusions}. — We introduced a unified algebraic framework for deriving the effective generator of coarse-graining time-dependent dynamical systems, covering both Hamiltonian methods (e.g., Lie-transform perturbation theory~\cite{cary_lie_1981}) and averaging-based approaches (e.g., time-coarse graining~\cite{gamel_time-averaged_2010}). We take an abstract perspective that views these approaches as specific examples of a general filtering transformation that removes irrelevant time scales. This allows us to relate the effective generator under this general averaging transformation to a generalized cumulant expansion~\cite{kubo_generalized_1962} in terms of the associated signature cumulants~\cite{bonnier_signature_2020}. This abstract perspective gives rise to a general formulation that applies to any time-dependent dynamical system with a linear generator, including Hamiltonian and non-Hamiltonian, conservative or dissipative, and quantum or classical systems. It also handles general time-dependencies, addressing non-adiabatic modulations, which are challenging with standard methods. Based on this formulation, we derived a simple procedure for calculating the effective generator for a simple convolution average. We then applied it to the damped Kapitza pendulum, demonstrating that our method surpasses state-of-the-art techniques by analyzing stability boundaries beyond existing literature. In addition, we demonstrate the ability of the method to treat non-adiabatic modulations by considering non-adiabatic modulation of the length of the pendulum, showing that this non-adiabaticity leads to effective dissipation, and then confirm our description by comparing it to an exact numerical calculation. These results establish a powerful tool for analyzing complex, strongly driven dynamics across physics and dynamical systems.

\bibliography{references,additional_ref}

\include{sm.tex}

\end{document}

%% file: preamble.tex
\usepackage[english]{babel}
\usepackage[utf8]{inputenc}
\usepackage[colorinlistoftodos, color=green!40, prependcaption]{todonotes}
\usepackage{soul}
\usepackage{amsthm}
\usepackage{mathtools}
\usepackage{physics}
\usepackage{xcolor}
\usepackage{graphicx}
\usepackage[left=23mm,right=13mm,top=35mm,columnsep=15pt]{geometry} 
\usepackage{adjustbox}
\usepackage{placeins}
\usepackage[T1]{fontenc}
\usepackage{lipsum}
\usepackage{csquotes}
\usepackage[pdftex, pdftitle={Article}, pdfauthor={Author}]{hyperref} 
\usepackage{simpler-wick}
\usepackage{listings}
\usepackage{nccmath}
\usepackage{comment}
\usepackage{multirow}
\usepackage{relsize}
\usepackage{wasysym}
\usepackage{natbib}
\usepackage{stmaryrd}
\usepackage{comment}

\definecolor{codegreen}{rgb}{0,0.6,0}
\definecolor{codegray}{rgb}{0.5,0.5,0.5}
\definecolor{codepurple}{rgb}{0.58,0,0.82}
\definecolor{backcolour}{rgb}
{0.95,0.95,0.92}
\definecolor{ForestGreen}{RGB}{34,139,34}

\lstdefinestyle{mystyle}{
    backgroundcolor=\color{backcolour},   
    commentstyle=\color{codegreen},
    keywordstyle=\color{magenta},
    numberstyle=\tiny\color{codegray},
    stringstyle=\color{codepurple},
    basicstyle=\ttfamily\footnotesize,
    breakatwhitespace=false,         
    breaklines=true,                 
    captionpos=b,                    
    keepspaces=true,                 
    numbers=left,                    
    numbersep=5pt,                  
    showspaces=false,                
    showstringspaces=false,
    showtabs=false,                  
    tabsize=2
}
\lstdefinelanguage{Julia}%
  {morekeywords={abstract,break,case,catch,const,continue,do,else,elseif,%
      end,export,false,for,function,immutable,import,importall,if,in,%
      macro,module,otherwise,quote,return,switch,true,try,type,typealias,%
      using,while},%
   sensitive=true,%
   alsoother={},%
   morecomment=[l]\#,%
   morecomment=[n]{\#=}{=\#},%
   morestring=[s]{"}{"},%
   morestring=[m]{'}{'},%
}[keywords,comments,strings]%

%% file: sm.tex
\appendix
\onecolumngrid
\section{Supplementary material for "Generalized time-coarse graining of Hamiltonian Systems"}

\subsection{Derivation of the generator of the averaged dynamics}
Our derivation assumes that the filtered dynamics evolves according to a linear generator, and can be written in the form of,
\begin{equation*}
    \dot{\overline{z}} = \mathcal{U}\overline{z}
\end{equation*}
We will show here that this form is a natural result of the linearity of the averaging transformation. Taking the time-derivative of $\overline{z} = T_wz$,
\begin{equation*}
    \dot{\overline{z}} = \dot{T}_w z + \overline{\dot{z}} 
\end{equation*}
using our original equation $\dot{z} = \mathcal{L}z$,
\begin{equation}
    \dot{\overline{z}} = \dot{T}_w z + T_w(\mathcal{L}z)
\end{equation}
Assuming the transformation is invertible $\overline{z} = T_w^{-1}z$,
\begin{equation}
\label{eq:generator_using_T}
    \dot{\overline{z}} = \bigg( \dot{T}_w T_w^{-1} + T_w \mathcal{L} T_w^{-1} \bigg) \overline{z}
\end{equation}
where we factored out $\overline{z}$ using the linearity of $T_w$. It is clear now that the dynamics are induced by a linear generator, given by a simple frame-change of the original generator. 
\begin{equation}
    \mathcal{U} = \dot{T}_w T_w^{-1} + T_w \mathcal{L} T_w^{-1} 
\end{equation}

\subsection{Derivation of the cumulant generating function using a Dyson expansion}
The propagator of the time-averaged dynamics is the average propagator,
\begin{equation}
    \overline{z}(t) = \overline{\mathcal{S}(t)}z_0 \equiv \mathcal{M}(t) z_0
\end{equation}
where for brevity we assumed $t_0 = 0$, and $z_0=z(0)$, and we defined the propagator of the time-averaged dynamics explicitly, $\mathcal{M}(t) \equiv \overline{\mathcal{S}(t)}$.  

The averaging procedure is chosen such that it simplifies the problem, for example by removing fast micro-motion. For time-coarse graining, this time-averaging operation is simply the convolution integral,
\begin{equation}
    \overline{\square} = \int_{-\infty}^{\infty} w(\tau) \square(t-\tau) d\tau    
\end{equation}
where $w(\tau)$ is a window function chosen to average over the fast micro-motion of the problem.

Our goal is to find an expression for the generator of $\overline{z}$, which would be simpler than the original generator, assuming an appropriate choice of the average.
\begin{equation}
    \frac{d\overline{z}}{dt} = \mathcal{U}(t) \overline{z}
\end{equation}
These definitions imply that $\mathcal{M}(t)$  is the time-ordered exponential of $\mathcal{U}(t)$, or equivalently,  that $\mathcal{U}(t)$ is the log-derivative of $\mathcal{M}(t)$,
\begin{subequations}
    \begin{equation}
        \mathcal{M}(t) = \exp_t \int_0^t dt' \mathcal{U}(t') 
    \end{equation}
    \begin{equation}
    	\frac{\rm d}{\rm dt}\mathcal{M}(t) = \mathcal{U}(t)\mathcal{M}(t)
\end{equation}
\end{subequations}
In order to find the effective generator $\mathcal{U}(t)$, we must invert the log-derivative equation. Since the averaging operation is not strictly invertible, generally it is not guaranteed that $\mathcal{M}(t)$ is invertible, and that we can find an explicit expression for the log-derivative $\mathcal{K}(t)$. However, it can be approximately inverted by expanding the operators in a formal power series.

We start by expanding the propagator in a Dyson series \cite{dyson_thesmatrix_1949},
\begin{subequations}
\begin{equation}
    \mathcal{M}(t) = \sum_{n=1}^\infty \varepsilon^n \mathcal{M}_n(t)
\end{equation}
\begin{equation}
    \mathcal{M}_n(t) = \overline{\mathcal{S}_n(t)} = \overline{\int_0^t dt_1 \int_0^{t_1} dt_2 \cdots \int_0^{t_{n-1}} dt_n \mathcal{L}(t_1) \cdots \mathcal{L}(t_n)}.
\end{equation}
\end{subequations}
So, for example,
\begin{align*}
    &\mathcal{M}_1(t) = 
    \overline{\int_0^t dt_1 \mathcal{L}(t_1)} \\
    &\mathcal{M}_2(t) = \overline{\int_0^t dt_1 \int_0^t dt_2 \mathcal{L}(t_1) \mathcal{L}(t_2)} \\
    &\vdots
\end{align*}

Similarly, we formally expand the generator in a formal power series as well,
\begin{equation}
        	\mathcal{U}(t;\varepsilon) = \sum_{n=1}^{\infty} \varepsilon^n \mathcal{U}_n(t)
\end{equation}
As originally defined by Kubo et. al., the different $\mathcal{M}_n$ can be thought of as generalized ordered moments, while the operator $\mathcal{U}_n(t)$ can be thought of as generalized cumulants, owing to their exponential relationship; similar expression also arise in the definition of the expected signature, in the study of rough paths. 

In principle, we have everything we need in order to calculate the different $\mathcal{M}_n(t)$.  Note that we are solving an inverse problem compared to the original one. In the original, we had the generator and tried to solve for propagator. Here we have the propagator $\mathcal{M}(t)$ (at least formally), and we try to solve for the generator $\mathcal{U}(t)$.  Plugging in the series expansions into the log-derivative and matching powers yields a recursive expression,
\begin{equation}
    	\dot{\mathcal{M}}_n(t) = \mathcal{U}_n(t) + \sum_{m=1}^{n-1} \mathcal{U}_m(t) \mathcal{M}_{n-m}(t) \rightarrow \boxed{\mathcal{U}_n(t) = \dot{\mathcal{M}}_n(t) - \sum_{m=1}^{n-1} \mathcal{U}_m(t) \mathcal{M}_{n-m}(t)}
\end{equation}
This recursion can be explicitly solved to give a closed form expression,
\begin{equation*}
    \mathcal{U}_n = \sum_{k=1}^n (-1)^{k+1} \sum_{(n_1,\ldots,n_k) \in \text{Comp}_k(n)} \dot{\mathcal{M}}_{n_1} \mathcal{M}_{n_2}\cdots \mathcal{M_{n_k}}
\end{equation*}
where $\text{Comp}_k(n)$ refers to the set of ordered composition of the integer $n$ into $k$ positive integers.
\begin{equation}
    	\text{Comp}_k(n) \equiv  \bigg\{ (n_1, n_2,\ldots, n_k) \in \mathbb{N}^k  \quad\bigg|\quad n_1 + n_2 + \cdots + n_k = n, \quad n_i \ge 1 \bigg\}
\end{equation}

It is helpful to include a couple of examples to explain the formulas and the calculations. Writing a few terms explicitly,
\begin{align*}
	\mathcal{U}_1 =& \dot{\mathcal{M}}_1 = \overline{\mathcal{L}(t)} \\
	\mathcal{U}_2 =&\dot{\mathcal{M}}_2 - \dot{\mathcal{M}}_1 \mathcal{M}_1 = \overline{\mathcal{L}(t) \int_0^t dt_1 \mathcal{L}(t_1)} - \overline{\mathcal{L}(t)} \overline{\int_0^t dt_1 \mathcal{L}(t_1)} \\
	\mathcal{U}_3 =& \dot{\mathcal{M}_3} - (\dot{\mathcal{M}}_1 \mathcal{M}_2 + \dot{\mathcal{M}}_2\mathcal{M}_1) + \dot{\mathcal{M}}_1 \mathcal{M}_1 \mathcal{M}_1 \\
		=& \overline{\mathcal{L}(t) \int_0^t dt_1 \mathcal{L}(t_1) \int_0^t dt_2 \mathcal{L}(t_2)} \\
		&- \bigg(\overline{\mathcal{L}(t)} \overline{\int_0^t dt_1 \mathcal{L}(t_1) \int_0^t dt_2 \mathcal{L}(t_2)} + \overline{\mathcal{L}(t) \int_0^t dt_1 \mathcal{L}(t_1)} \cdot \overline{\int_0^t dt_2 \mathcal{L}(t_2)} \bigg) \\
		&+ \overline{\mathcal{L}(t)} \overline{\int_0^t dt_1 \mathcal{L}(t_1)} \overline{\int_0^t dt_2 \mathcal{L}(t_2)}
\end{align*}

\subsection{Derivation of the cumulant-generating function using a Magnus expansion}
An alternative way to invert the log-derivative equation is by expanding the propagator in a Magnus expansion  \cite{magnus_exponential_1954}, where we now define the propagator as an \textbf{ordinary} exponential, 
\begin{equation}
    \exp \mathcal{K}(t) = \mathcal{M}(t)
\end{equation}
This is very similar to the definition we used in the previous derivation, except that here the exponential is \textbf{unordered}, and $\mathcal{K}(t)$ is not the direct log-derivative and does not generate the dynamics. This Magnus-type derivation is also particularly useful for problems with a Hamiltonian structure and finite algebras, and in principle can have better convergence properties.

This can also be seen as an alternative definition for the cumulant-generating function -- since we now deal with operators rather than variables, the definition of the exponential and the cumulant-generating function depends on the ordering.

Similar to the previous derivation, we can find an explicit expression for $\mathcal{K}(t)$ by expanding it and $\mathcal{M}(t)$ in a formal power series,
\begin{align}
	&\mathcal{M}(t) = \mathcal{I} + \sum_{n=1}^\infty \mathcal{M}_n(t) = \exp \mathcal{K}(t) \\
	&\mathcal{K}(t) = \sum_{n=1}^\infty \mathcal{K}_n(t) = \log \mathcal{M}(t) 
\end{align}

Using the expansion, $\log(\mathcal{I} + X) = \sum_{k=1}^\infty \frac{(-1)^{k+1}}{k} X^k$,
\begin{equation}
    \sum_{n=1}^{\infty}\mathcal{K}_n = \sum_{k=1}^\infty \frac{(-1)^{k+1}}{k} \left( \sum_{l=1}^\infty \mathcal{M}_l(t)\right)^k = \sum_{k=1}^\infty \frac{(-1)^{k+1}}{k} \sum_{n_i\ge1}^\infty \mathcal{M}_{n_1}\cdots\mathcal{M}_{n_k}    
\end{equation}
	
By grouping terms of matching powers,
\begin{equation}
\mathcal{K}_n = \sum_{k=1}^n \frac{(-1)^{k+1}}{k} \sum_{(n_1, \ldots, n_k) \in\text{Comp}_k(n)} \mathcal{M}_{n_1}\cdots\mathcal{M}_{n_k}
\end{equation}
where $\mathcal{M}_n$ are the ordered moments and $\text{Comp}_k (n)$ is the set of compositions $k$ summing up to $n$, as defined above.

To make the formulas clearer, we write a few terms explicitly,
\begin{align}
	&\mathcal{K}_1 = \mathcal{M}_1 = \overline{\int_0^t dt_1 \mathcal{L}(t_1)} \\
	&\mathcal{K}_2 = \mathcal{M}_2 - \frac{1}{2} \mathcal{M}_1^2 = \overline{\int_0^t dt_1 \int_0^t dt_2 \mathcal{L}(t_1) \mathcal{L}(t_2)} -  \overline{\int_0^t dt_1 \mathcal{L}(t_1)} \cdot \overline{\int_0^t dt_2 \mathcal{L}(t_2)} \\
	&\mathcal{K}_3 = \mathcal{M}_3 - \frac{1}{2}(\mathcal{M}_1 \mathcal{M}_2 + \mathcal{M}_2 \mathcal{M}_1) + \frac{1}{3} \mathcal{M}_1^3 \\
	&\vdots
\end{align}

However, $\mathcal{K}(t)$ is still not the generator of the dynamics. In order to find the generator $\mathcal{U}(t)$., we need to solve the log-derivative equation. Plugging in $\mathcal{M}(t) = \exp \mathcal{K}(t)$,
\begin{equation}
    \frac{\rm d}{\rm dt}\exp \mathcal{K}(t) = \mathcal{U}(t)\exp \mathcal{K}(t)
\end{equation}
Since the exponential is not time-ordered, the left-hand side of the equation is now given by the Wilcox formula for the exponential map,
\begin{equation}
    \frac{d}{dt} \exp \mathcal{K}(t) = \int_0^1 e^{\alpha \mathcal{K}(t)} \dot{\mathcal{K}} e^{-\alpha \mathcal{K}(t)} d\alpha \cdot \exp \mathcal{K}(t)
\end{equation}
which immediately yields,
\begin{equation}
    \mathcal{U}(t) = \int_0^1 e^{\alpha \mathcal{K}(t)} \dot{\mathcal{K}} e^{-\alpha \mathcal{K}(t)} d\alpha  
\end{equation}
	
To simplify further, we use the common identity for the exponential adjoint,
\begin{equation}
    e^{X}Ye^{-X} = \sum_{n=0}^\infty \frac{s^n}{n!} \text{ad}_X^n Y
\end{equation}
Now, integrating term by term, we find,
\begin{equation}
    \mathcal{U}(t) = \sum_{n=0}^{\infty} \frac{1}{(n+1)!} \text{ad}_{\mathcal{K}}^n \dot{\mathcal{K}}
\end{equation}
where we defined the repeated commutator,
\begin{align}
	&\text{ad}^0_{A} B = B \\
	&\text{ad}^1_{A} B = [A, B] = AB-BA \\
	&\text{ad}^{n+1}_{A} = [A, \text{ad}_A^n B]
\end{align}
This gives us an explicit series expansion for the generator $\mathcal{U}(t)$ in terms of the unordered cumulant-generating function $\mathcal{K}(t)$, where different terms in the sum capture the effects of the non-commutativity of the terms.
\begin{subequations}
\begin{equation}
    \mathcal{U}(t) = \sum_{n=0}^\infty \mathcal{U}^{(n)}(t)
\end{equation}
\begin{equation}
		\mathcal{U}^{(n)}(t) = \frac{1}{(n+1)!} \text{ad}_{\mathcal{K}}^n \dot{\mathcal{K}}
\end{equation}
\end{subequations}

This grading is useful for finite algebras where the generators close, and systematically shows the difference between the generalized cumulants and the ordinary cumulants, arising from the non-commutativity of the operators; In the commutative limit, only $\dot{\mathcal{U}}(t)$ contributes to the generator $\mathcal{K}(t)$, and the terms in its expansion become the classical cumulant rates. This also implies that the generator (the signature cumulant) lie in the associated Lie algebra of the problem. It is only because of our truncation of terms that the generator becomes non-Hamiltonian.

\subsection{LPT as a filtering transformation}
Often in physics, we deal with Hamiltonian systems,
\begin{equation}
    \dot{z} = \mathcal{L}_{H(t)} z
\end{equation}
where $\mathcal{L}_{H(t)}$ is the Lie-derivative with respect to the Hamiltonian $H(t)$.

In \textit{Lie-perturbation theory} (LPT) \cite{dewar_renormalised_1976, cary_lie_1981, dragt_lie_1976, venkatraman_static_2022}, we simplify the dynamics by seeking canonical transformations that eliminate rapidly oscillating motion,
\begin{subequations}
\begin{align}
    &T_w = \mathcal{S}_{w(t;\varepsilon)} =  \exp_\varepsilon \left( \int_0^\varepsilon \mathcal{L}_{w(t;\varepsilon')} d\varepsilon' \right) 
\end{align}
\end{subequations}
where we defined an auxiliary ``time'' variable $\varepsilon$, and ordered the exponential with respect to it. This transformation maps phase-space coordinates into action-angle variables, where the dynamics under a simplified Hamiltonian $K$ (often called the Kamiltonian or gyrocenter Hamiltonian in plasma physics) with $\mathcal{U} = \mathcal{L}_K$. 

We can easily show that using Eq.\ref{eq:generator_using_T},
\begin{equation}
    \mathcal{U} = (\partial_t\mathcal{S}_w)\mathcal{S}_{w}^{-1} + \mathcal{S}_w \mathcal{L}_H \mathcal{S}_w^{-1}
\end{equation}

Using $\Phi \mathcal{L}_H \Phi^{-1} = \mathcal{L}_{\Phi[H]}$, we may write,
\begin{equation}
     \mathcal{S}_w \mathcal{L}_H \mathcal{S}_w^{-1} = \mathcal{L}_{S_W H}
\end{equation}

Using the Wilcox formula, we may write,
\begin{equation}
    (\partial_t\mathcal{S}_w)\mathcal{S}_w^{-1} = \int_0^1 \mathcal{S}_w(\varepsilon, \varepsilon') (\partial_t \mathcal{L}_w) \mathcal{S}_w^{-1}(\varepsilon,\varepsilon') d\varepsilon' \cdot \mathcal{S}_w  \mathcal{S}_w^{-1} = \mathcal{L}_{\int_0^\varepsilon \mathcal{S}_w(\varepsilon, \varepsilon')\dot{w} d\varepsilon'}
\end{equation}
where we used the linearity of the Lie derivative and the transformation rule to put everything under the Lie-derivative. Overall, this yields an expression for the generator of the dynamics in the new frame,
\begin{subequations}
 \begin{align}
    &\mathcal{U} = \mathcal{L}_K \\
    &K = \int_0^1\mathcal{S}_w(\varepsilon, \varepsilon')\dot{w} d\varepsilon' + \mathcal{S}_w(\epsilon,0) H
\end{align}   
\end{subequations}
where it is clear now that the dynamics is generated by a Lie-derivative with respect to a Hamiltonian-like function $K$, known in the literature as the Kamiltonian.

For integrable systems, such symplectic transformations naturally yield a simple description. However, since integrability is rare, $K$ and $w$ are found by iteratively, using a perturbative expansion of in $\varepsilon$. 

Although the simple Hamiltonian structure of the generator is attractive, the complex structure of the symplectic transformation, and the fact that it has to be expanded in a series, complicates the formulas of LPT. 

\subsection{TCG and the convolution average}
\textit{Time coarse-graining} (TCG)~\cite{gamel_time-averaged_2010, cresser_coarse-graining_2017, lee_effective_2018, majenz_coarse_2013, bello_systematic_2024}, developed in the context of quantum optics, departs from traditional Hamiltonian techniques by replacing canonical transformations with a simple convolutional averaging:
\begin{equation}
    T_w = E_w \equiv \int_{-\infty}^{\infty} w(\tau)\, \square(t-\tau)\, d\tau.
\end{equation}
Here, the filter function $p(t)$ acts as a linear time-invariant filter—typically a low-pass or band-pass filter—selecting a specific frequency band of the system's dynamics. Unlike canonical transformations, the convolution average is time-independent (i.e., \( \dot{T}_w = 0 \), or equivalently, \( [E_w, \partial_t] = 0 \)), which significantly simplifies the algebraic structure of the theory.

The choice of the filter is guided by physical intuition and the characteristic time-scales of the system. For instance, if the motion is predominantly harmonic around a natural frequency \( \omega_0 \), a band-pass filter centered at \( \omega_0 \) is a natural choice. However, convolution filters are not universally optimal. In non-stationary problems, where the spectral content varies with time, a fixed filter may fail to capture essential dynamics. If the filter is poorly matched to the system, the resulting coarse-grained description may poorly approximate the true evolution. In some instances, the filter may be chosen to model an experimental apparatus, in which case the divergence from the true evolution of the system is a feature that describes the loss of information due to the incomplete measurement \cite{bello_systematic_2025}.

An important theoretical consideration is the invertibility of the convolution. Convolution filters are not strictly invertible unless they have infinite spectral support. Any truncation of non-zero spectral components renders the transformation non-invertible. While this might appear to threaten the consistency of the theory, it is important to remember that TCG is a perturbative framework. In practice, the transformation need only be invertible up to order \( \mathcal{O}(\varepsilon^n) \). In other words, as long as spectral components smaller than the truncation error are the only ones removed, no inconsistency is introduced at the given perturbative order.

Neglecting of small, non-zero terms, allows for the simplification of the generator. However, this simplification comes at a cost: since the transformation is only approximately invertible, the generator is generally non-Hamiltonian. As a result, the effective dynamics may include non-Lie terms. In quantum mechanics—and within the standard TCG framework—this leads naturally to a Lindblad-like evolution, characteristic of open quantum systems and associated with decoherence or information loss. From the perspective of Lie perturbation theory (LPT), one may interpret the original system as comprising multiple weakly coupled subsystems with different time-scales. In this view, TCG corresponds to averaging over the fast subsystem, inducing effective non-Hamiltonian terms due to the residual coupling between time-scales.

\subsubsection{Properties of the convolution average}
The convolution average is very simple compared to symplectic transformation -- it is linear, time-independent, and has a very simple algebraic structure. This makes the derivation and formulas of TCG simpler than those of LPT, at the cost of a more complicated effective generator.

The convolution average is linear and symmetric,
\begin{subequations}
\begin{align}
    E_{\lambda w+ \mu v} = \lambda E_w + \mu E_v \\
    E_w v = E_v w
\end{align}    
\end{subequations}

In Frequency-domain, the convolution operators are diagonalized, and become simple multiplication.
\begin{equation}
    \hat{E}_W = \mathcal{F} E_W \mathcal{F}^{-1} = \mathcal{F} \left[ W*\mathcal{F}^{-1} \right] = \hat{W}
\end{equation}
where $\mathcal{F}$ denotes the Fourier transform operator, $\hat{\square}$ denotes the transformed function/operator, and $*$ denotes convolution.

The convolution operators are easily composed. For two convolution operators,
\begin{equation}
    E_w E_v = E_{w*z}
\end{equation}
In frequency domain this simplifies to a simple product.
\begin{equation}
    \hat{E}_{\hat{w}} \hat{E}_{\hat{v}} = \hat{E}_{\hat{w} \hat{v}}
\end{equation}

Convolution with a delta function is simply the identity, making the algebra of convolution operators unital,
\begin{equation}
    E_{\delta(t)} = \hat{E}_1 = I
\end{equation}
The most important property of convolution operators is that they can approximately invertible. Using the composition property, for two convolution operators to be inverses,
\begin{equation}
    E_w E_v = E_{w*v} = I \rightarrow (w*v)(t) =\delta(t)
\end{equation}

This is more explicit in frequency-domain,
\begin{equation}
    \hat{E}_w \hat{E}_v = \hat{E}_{\hat{w}\hat{v}} = I \rightarrow \hat{w} \hat{v} = 1
\end{equation}

This implies,
\begin{equation}
    \hat{E}_{\hat{w}}^{-1} = \hat{E}_{\hat{w}^{-1}} \qquad \hat{w}[\omega] \neq 0
\end{equation}
i.e. the inverse of a convolution operator is only defined if its window function $\hat{w}[\omega]$ has infinite support, which is not physical nor useful. 

This may seem to imply a contradiction with our requirement of the transformation $T_w$, and indeed it does. The saving grace of the TCG approach is that it is a perturbative method. The derived generators need only be approximately invertible up to the expansion error $O(\varepsilon^n)$.

\subsection{Harmonic Time-dependencies}
In many problems of interest, the external time-dependence is harmonic with $f_\omega(t) = a_\omega e^{-i\omega t}$. Harmonic time-dependencies are easy to work with, since they are closed under multiplication, integration, and differentiation (i.e. under these operations, a harmonic function remains a harmonic function). This makes the ordered moments much easier to work with, since the nested integral now yields a simple closed-form formula:
\begin{subequations}
  \begin{equation}
    M_\omega = \frac{i^n}{\omega_{\uparrow_n}} \overline{e^{-i(\omega_1 + \cdots \omega_n)t}}
\end{equation}
\begin{equation}
    \dot{M}_\omega = -i(\omega_1 + \cdots + \omega_n) M_\omega
\end{equation}
\end{subequations}
where we define the function $\omega_{\uparrow_n}$ is the product of tail-cumulative partial sums of the vector,
\begin{equation}
    \omega_{\uparrow_n} = \prod_{k=1}^n \left( \sum_{i=k}^n \omega_k \right) = (\omega_1 + \cdots + \omega_n)(\omega_2 + \cdots + \omega_n)\cdots(\omega_{n-1} + \omega_n)\omega_n 
\end{equation}
we choose the up-arrow symbol $\uparrow_n$ to indicate that the lower index increases between the partial sums.

\subsection{Worked Example I -- the Kapitza pendulum}

\subsubsection{The Hamiltonian and dissipator formulation of the damped Kapitza pendulum}

As discussed in the main text, we consider a classical damped Kapitza pendulum described by the Hamiltonian,
\begin{equation}
\begin{split}
H(\phi,p_{\phi},t)
=
\frac{p_{\phi}^{2}}{2m l^{2}}
-
mg l \left( 1- \lambda \frac{\nu^2}{\omega_0^2}   \cos(\nu t) \right) \cos\phi.	
\end{split}
\end{equation}
where $\omega_{0} \equiv \sqrt{\frac{g}{l}}$ and $\lambda$ is the dimensionless drive strength.
We model the damping in the problem using the double-bracket formulation, using the following dissipator,
\begin{equation}
\begin{split}
D(\phi,p_{\phi})
=
-
\frac{1}{2} Q^{-1} \omega_{0} p_{\phi}^{2}
\end{split}
\end{equation}
where $Q$ is the quality factor of the resonator.

The hallmark feature of the Kapitza pendulum is the emergence of dynamically stable points -- when the modulation of the pendulum crosses a certain threshold, the pendulum becomes stable when it is in an inverted position, upside-down. To study the stability around this inverted position, it is more convenient to work with our coordinates centered around the inverted position,
\begin{equation}
\begin{split}
\left( \theta, p_{\theta} \right)
\equiv
\left( \pi - \phi, - p_{\phi} \right)
\end{split}
\end{equation}
in this frame the Hamiltonian and dissipator become,
\begin{equation}
\begin{split}
H(\theta,p_{\theta},t)
=
\frac{p_{\theta}^{2}}{2m l^{2}}
+
m gl \left( 1- \lambda  \frac{\nu^{2}}{\omega_0^2} \cos \nu t \right) \cos\theta		
\end{split}
\end{equation}
\begin{equation}
\begin{split}
D(\theta,p_{\theta})
=
-
\frac{1}{2} Q^{-1} \omega_{0} p_{\theta}^{2}
\end{split}
\end{equation}
In this formulation, the equations of motion can be written in terms of the Poisson-bracket and the double-bracket,
\begin{equation}
\begin{split}
&
\partial_{t} \theta
=
\left\{ \theta, H \right\}
+
\llbracket \theta, D \rrbracket
\equiv
\frac{\partial \theta }{\partial \theta} \frac{\partial H }{\partial p_{\theta}}
-
\frac{\partial \theta }{\partial p_{\theta}} \frac{\partial H }{\partial \theta}
+
\frac{\partial \theta }{\partial p_{\theta}} \frac{\partial D}{\partial p_{\theta}}
=
\frac{ p_{\theta} }{ ml^{2} }\\
&
\partial_{t} p_{\theta}
=
\left\{ p_{\theta}, H \right\}
+
\llbracket p_{\theta}, D \rrbracket
\equiv
\frac{\partial p_{\theta} }{\partial \theta} \frac{\partial H }{\partial p_{\theta}}
-
\frac{\partial p_{\theta} }{\partial p_{\theta}} \frac{\partial H }{\partial \theta}
+
\frac{\partial p_{\theta} }{\partial p_{\theta}} \frac{\partial D}{\partial p_{\theta}}
=
-
Q^{-1} \omega_{0} p_{\theta}
+
mgl \big( 1 +  \lambda \frac{\nu^{2}}{\omega_0^2} \cos\nu t \big) \sin\theta
\end{split}
\end{equation}
which yields the following second-order equation for $\theta$,
\begin{equation}
\begin{split}
\partial_{t}^{2} \theta
+
Q^{-1} \omega_{0} \partial_{t} \theta
-
\omega_{0}^{2} \left( 1 - \lambda \frac{\nu^{2}}{\omega_{0}^{2}} \cos\nu t \right) \sin\theta
=
0
.
\end{split}
\end{equation}
which is the equation of a damped harmonic oscillator, whose resonant frequency is modulated at frequency $\nu$.

For bookkeeping purposes, it is illuminating to use a dimensionless presentation, in units of the modulation frequency $\tilde{t} = \nu t$. This motivates defining the dimensionless momentum,
\begin{equation}
    \tilde{p}_\theta = \frac{p_\theta}{ml^2 \nu}
\end{equation}
as used in the main text. We normalize the Hamiltonian and dissipator by the modulation energy $ml^2 \nu^2$ and substitute in the dimensionless momentum $\tilde{p}_\theta$ to obtain the dimensionless Hamiltonian,
\begin{subequations}
\begin{equation}
    \tilde{H}(\theta, \tilde{p}_\theta, \tilde{t}) = \frac{\tilde{p}_\theta^2}{2} + \Gamma^2\left(1 - \Lambda \cos \tilde{t} \right) \cos \theta
\end{equation}
\begin{equation}
    \tilde{D}(\theta, \tilde{p}_\theta) = -\frac{1}{2} Q^{-1} \Gamma \tilde{p}_\theta^2
\end{equation}
\end{subequations}
where $\Lambda$ is the modulation strength under the new normalization and we defined the dimensionless frequency, 
\begin{equation}
    \Gamma = \omega_0/\nu    
\end{equation}
which is just the natural frequency of the pendulum $\omega_0$ in units of the modulation frequency.

We can split our Hamiltonian into a time-independent base, and a time-dependent perturbation.
\begin{equation}
    \tilde{H}(\theta, \tilde{p}_\theta, \tilde{t}) = \underbrace{\frac{\tilde{p}_\theta^2}{2} + \Gamma^2 \cos \theta}_{\tilde{H}_0(\theta, \tilde{p}_\theta)} \,\, \underbrace{-\Lambda \Gamma^2 \cos \tilde{t} \cos \theta}_{H_1(\theta, \tilde{t})}
\end{equation}
This defines our perturbation parameter $\epsilon = \Lambda \Gamma^2 \ll 1$, and the regimes where a perturbative treatment is justified -- weak drives and slow modulations, or fast modulations and strong drives.

\subsubsection{Stabilization of the inverted position}

As suggested in \cite{Blackburn_1992AmJPh}, the steady-state value of $\theta$ can be at the fully inverted position $\theta = 0$ when the non-normalized modulation strength $\lambda$ is within a certain stability range.
In this dynamically stable regime, the dynamics around the inverted steady-state can be assumed to be slow and well separated from the other timescales of the problem, which makes it perfectly suitable for analysis using our time-coarse graining approach. More precisely, in this regime the angle of pendulum is slowly-varying and can be treated as an averaged quantity $\overline{\theta} \ll 1$, as long as the coarse graining time scale $\tau$ is much greater than the modulation time-scale $\nu^{-1}$. 

Following our perturbative prescription, we can write the generator $\mathcal{L}$ as a Fourier series,
\begin{equation}
\begin{split}
\mathcal{L}
=
\mathcal{L}_{0}
+
e^{i \nu t} \mathcal{L}_{\nu}
+
e^{-i \nu t} \mathcal{L}_{-\nu}
\end{split}
\end{equation}

\begin{equation}
\begin{split}
\mathcal{L}_{0}
\equiv
\ \tilde{p}_\theta \partial_{\theta}
+
\big(
\Gamma^2 \sin\theta 
-
Q^{-1} \Gamma p_{\theta}
\big) \partial_{\tilde{p}_{\theta}}
\end{split}
\end{equation}
\begin{equation}
\begin{split}
\mathcal{L}_{\nu}
=
\mathcal{L}_{-\nu}
\equiv
-
\frac{\lambda}{2} \sin\theta \partial_{\tilde{p}_{\theta}}
.
\end{split}
\end{equation}
Note that these are the generators with respect to the dimensionless time $\tilde{t} = \nu t$. In order to recover units, we simply multiply the expressions by $\nu$.

The form of the generator is precisely the form we consider in the main text, with a harmonic time-dependence $f_\omega(t) = e^{-i\omega t}$. These terms define a three-letter "alphabet" from which the corrections to the effective generator would be formed.

Our approach for finding the stable points of the system is to perturbatively derive the generator and find its stable points.
It is easy then to derive the first-order effective generator:
\begin{equation}
    \mathcal{U}^{(1)}
    =
    \mathcal{L}_{0}
    =
    \left[
    \tilde{p}_{\theta} \partial_{\theta}
    +
    \left(
    \Gamma^2 \sin\theta
    -
    Q^{-1} \Gamma \tilde{p}_{\theta}
    \right)\partial_{\tilde{p}_{\theta}}
    \right] 
\end{equation}

The first-order results are analogous to naive averaging of the generator, which is equivalent to the well-known rotating-wave approximation. However, the time-averaged generator is simply $\mathcal{L}_{0}$, which predicts that the inverted position at $\theta = 0$ is unstable. 

Indeed, to recover the dynamically stable points of the problem, we must calculate the effective generator to higher orders in the perturbation. By symmetry, all even-order terms in the generator vanish. The next non-vanishing correction arises from the third-order perturbation.
\begin{equation}
    \mathcal{U}^{(3)}
=
-
\frac{\lambda^{2}}{2}
\left[
\frac{1}{2} \sin(2\theta) \partial_{\tilde{p}_{\theta}}
+
\sin\theta \left(\tilde{p}_{\theta} \cos\theta + \Gamma Q^{-1} \sin\theta \right) \partial_{\tilde{p}_{\theta}}^{2}
-
\sin^{2}\theta \cdot \partial_{\theta} \partial_{\tilde{p}_{\theta}}
\right] 
\end{equation}
As per our procedure, we calculate all three-letter words and their associated weights (given by the cumulants), and then truncate terms that are smaller than the perturbation order. At this order, we already observe the stabilization of the inverted position. To surpass the current state-of-the-art, we go beyond that. Here, we explicitly provide the fifth-order term of the effective generator.
\begin{align}
    \mathcal{U}^{(5)}
    =
    -
    \frac{\lambda^{2}}{2} 
    \Big[
    &
    \big(
    \left( 3\tilde{p}_{\theta}^{2} - Q^{-2} \Gamma^2 \right) \cos\theta
    -
    4\Gamma^2 \cos(2\theta)
    \big) \sin\theta \cdot \partial_{\tilde{p}_{\theta}}
    -
    6 \sin^{2}\theta \cdot  \tilde{p}_{\theta} \partial_{\theta}\\
    &
    -
    \left( 3\tilde{p}_{\theta}^{2} - Q^{-2}\Gamma^{2} - 4\Gamma^2\cos\theta \right) \sin^{2}\theta \cdot \partial_{\theta} \partial_{\tilde{p}_{\theta}}\\
    &
    -
    \sin\theta \left(
    4 \Gamma^2 \cos(2\theta) \tilde{p}_{\theta}
    +
    Q^{-3} \Gamma^3 \sin\theta
    -
    \cos\theta ( \tilde{p}_{\theta}^{3} - Q^{-2}\Gamma^2 \tilde{p}_{\theta} - 4 Q^{-1} \Gamma^3 \sin\theta )
    \right) \partial_{\tilde{p}_{\theta}}^{2}\\
    &
    +
    \frac{\lambda^{2}}{2} \cos\theta\sin^{3}\theta \cdot \partial_{\tilde{p}_{\theta}}^{3}
    +
    \frac{\lambda^{2}}{8} \sin^{4}\theta \cdot \partial_{\theta}\partial_{\tilde{p}_{\theta}}^{3}
    -
    \frac{\lambda^{2}}{8} \sin^{3}\theta \left( \tilde{p}_{\theta} 
\cos\theta + Q^{-1} \Gamma \sin\theta \right) \partial_{\tilde{p}_{\theta}}^{4}
    \Big]
\end{align}
We also calculate the 7th order contribution, but the expressions become convoluted and not useful, and we therefore omit them.

The stabilization of the inverted position is achieved when $\left( \theta, \tilde{p}_{\theta} \right) = \left( 0,0 \right)$ is a stable fixed point of the effective generator $\mathcal{U} = \mathcal{U}^{(1)} + \mathcal{U}^{(2)} + \mathcal{U}^{(3)} + \cdots$. Calculating the generator up to orders 3, 5 and 7, we obtain the following minimal stabilizing amplitudes:
\begin{equation}
\begin{split}
&\textrm{order 3}:
\big( \lambda_{0}^{(3)} \big)^{2}
=
2\Gamma^2\\
&\textrm{order 5}:
\big( \lambda_{0}^{(5)} \big)^{2}
=
\frac{2\Gamma^2}{1-4\Gamma^2-\Gamma^2 Q^{-2}}
\approx
2\Gamma^2 \left( 1 + 4\Gamma^2 + Q^{-2}\Gamma^2 \right)
\\
&\textrm{order 7}:
\big( \lambda_{0}^{(7)} \big)^{2}
=
\frac{8}{25}\bigg(
\sqrt{\frac{25 \Gamma^2}{2} + (1-Q^{2} \Gamma^2 + Q^{-4} \Gamma^4-4\Gamma^2+8 Q^{-2} \Gamma^4 + 16\Gamma^{4})^{2}} \\
&\qquad\qquad\qquad\quad
-
1
+
Q^{-2} \Gamma^2
-
Q^{-4} \Gamma^4
+
4\Gamma^2
-
8 Q^{-2} \Gamma^4 
-
16\Gamma^{4}
\bigg)\\
&\qquad\qquad\qquad\quad
\approx
2\Gamma^2 \left(
1
+
Q^{-2} \Gamma^2 
+
\frac{7\Gamma^2 - 75 Q^{-2} \Gamma^4 - 171 Q^{-4} \Gamma^6 }{8}
-
\frac{575\Gamma^{4} + 811 Q^{-2} \Gamma^6  -5578 Q^{-4}\Gamma^8}{32}
\right)
\end{split}
\end{equation}
As we discuss in the main text, this goes beyond the results known in the literature, adding additional dissipation-dependent corrections that are difficult to find using standard methods.

\subsubsection{Upper boundary of dynamical stabilization}

\begin{figure}[h!]
    \centering
\includegraphics[width=0.55\linewidth]{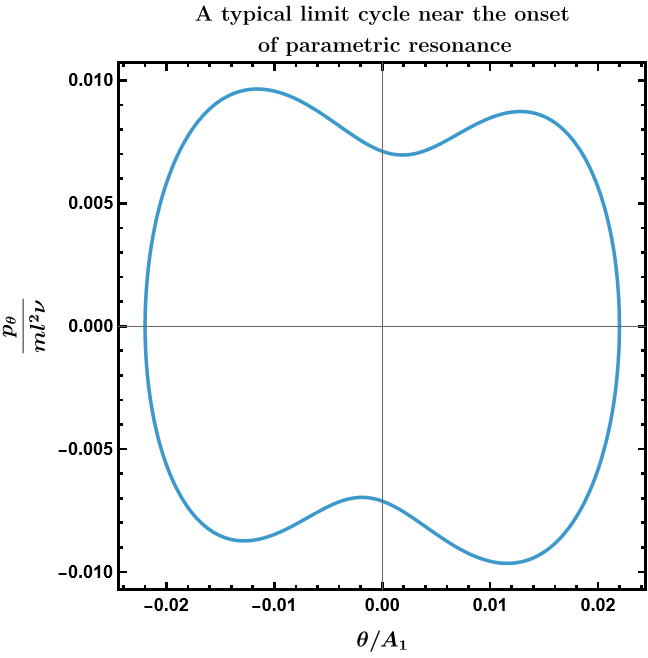}
    \caption{As the drive strength $\lambda$ is ramped up, limit cycles first emerge due to the parametric resonance at side-band frequencies $\nu/2$ and $3\nu/2$. The limit cycles are not symmetrical in general with nonzero $\alpha_{1}$ and $\alpha_{2}$ which comes from the dissipative dynamics.}
    \label{fig:Kapitza_parametric}
\end{figure}

As the modulation strength $\lambda$ is increased, it may cross a threshold where the dynamically stabilized fixed point breaks into a limit cycle due to the onset of parametric resonance. According to the literature~\cite{Blackburn_1992AmJPh, Butikov2001}, the limit cycles near the onset of parametric resonance can be (approximately) parametrized as a harmonic motion with two frequencies,
\begin{equation}
\label{Eq: limit cycle parameterization}
\begin{split}
&
\theta(t)
=
A_{1} \cos\left( \frac{\nu}{2} t + \alpha_{1} \right)
+
A_{3} \cos\left( \frac{3\nu}{2} t + \alpha_{3} \right)\\
&
\tilde{p}_{\theta}(t)
=
-
\frac{1}{2} A_{1} \sin\left( \frac{\nu}{2} t + \alpha_{1} \right)
-
\frac{3}{2} A_{3} \sin\left( \frac{3\nu}{2} t + \alpha_{3} \right)
\end{split}
\end{equation}
where $A_{1}, A_{3}, \alpha_{1}, \alpha_{3}$ are real-valued parameters. Such typical limit cycle is illustrated in Fig.\ref{fig:Kapitza_parametric}. 

These limit cycles are fast, on the order of the modulation frequency. In particular, one full cycle takes $4\pi / \nu$. This means that in the lab-frame, their dynamics is not well-separated from the fast time-scales in the problem, precluding a direct approach like the one we used in order to derive the lower stability threshold.

In order to study a particular limit cycle within our framework,
we need to transform to a frame rotating at the frequencies of the limit cycle and the corresponding sidebands. In those frames, the limit-cycle becomes slow, and its onset is signaled by the stable position of the canonically transformed phase-space coordinates $\left( \theta^{\textrm{CT}}, p_{\theta}^{\textrm{CT}} \right)$ becoming nonzero. For example, suppose that a stable limit cycle is indeed parametrized as in Eq.(\ref{Eq: limit cycle parameterization}), then one may consider the following two time-dependent canonical transformations (CT), comprised of displacement and rotation:

\begin{itemize}
    \item CT1: Displace $\left(\theta, p_{\theta}\right)$ by $\left( A_{1}\cos\left(\frac{\nu t}{2}\right), - \frac{\nu}{2} m l^{2} A_{1} \sin\left( \frac{\nu t}{2} \right) \right)$ and then rotate in the counter-clockwise direction at rate $\frac{3\nu t}{2}$.
    \item CT2: Displace $\left(\theta, p_{\theta}\right)$ by $\left( A_{3}\cos\left(\frac{3\nu t}{2}\right), - \frac{3\nu}{2} m l^{2} A_{3} \sin\left( \frac{3\nu t}{2} \right) \right)$ and then rotate in the counter-clockwise direction at rate $\frac{\nu t}{2}$.
\end{itemize}
In other words, we displace one side-band and rotate the other, and the two transformation swap which side-band is displaced and which one is rotated. It is convenient to define the following phase-space coordinates in the two rotating frames:
\begin{equation}
\begin{cases*}
&
$
\theta^{\textrm{CT1}}
=
\cos\left( \frac{3\nu}{2} t \right)
\cdot\left[ \theta - A_{1} \cos\left( \frac{\nu t}{2} \right) \right]
-
\frac{\sin\left( \frac{3\nu}{2} t \right)}{ml^{2} \frac{3\nu}{2}}
\cdot\left[ p_{\theta} - ml^{2} \left( -\frac{\nu}{2} A_{1} \sin\left(\frac{\nu}{2} t\right) \right) \right]
$\\
&
$
\tilde{p}_{\theta}^{\textrm{CT1}}
=
\sin\left( \frac{3\nu}{2} t \right)
\cdot\left[
\theta
-
A_{1} \cos\left( \frac{\nu}{2} t \right)
\right]
+
\frac{\cos\left( \frac{3\nu}{2} t \right)}{ml^{2} \frac{3\nu}{2}}
\cdot\Big[
p_{\theta}
-
ml^{2}\left(
-
\frac{\nu}{2} A_{1} \sin\left( \frac{\nu}{2} t \right)
\right)
\Big]
$
\end{cases*}
\end{equation}
\begin{equation}
\begin{cases*}
&
$
\theta^{\textrm{CT2}}
=
\cos\left( \frac{\nu}{2} t \right)
\cdot\left[ \theta - A_{3} \cos\left( \frac{3\nu t}{2} \right) \right]
-
\frac{\sin\left( \frac{\nu}{2} t \right)}{ml^{2} \frac{\nu}{2}}
\cdot\left[ p_{\theta} - ml^{2} \left( -\frac{3\nu}{2} A_{3} \sin\left(\frac{3\nu}{2} t\right) \right) \right]
$\\
&
$
\tilde{p}_{\theta}^{\textrm{CT2}}
=
\sin\left( \frac{\nu}{2} t \right)
\cdot\left[
\theta
-
A_{3} \cos\left( \frac{3\nu}{2} t \right)
\right]
+
\frac{\cos\left( \frac{\nu}{2} t \right)}{ml^{2} \frac{\nu}{2}}
\cdot\Big[
p_{\theta}
-
ml^{2}\left(
-
\frac{3\nu}{2} A_{3} \sin\left( \frac{3\nu}{2} t \right)
\right)
\Big]
$
\end{cases*}
\end{equation}

In both of the frames the limit-cycle is mapped to a central point that is parametrized by the phase and amplitude of one of the side-bands of the limit-cycle. Under CT1, the limit cycle around $\left( \theta, p_{\theta} \right) = \left( 0, 0 \right)$ with sideband frequencies $\nu/2$ and $3\nu/2$ is mapped to the point,
\begin{equation}
\label{Eq: center point after CT1}
\begin{split}
\left( \theta^{\textrm{CT1}}, \tilde{p}_{\theta}^{\textrm{CT1}} \right)
=
Z_{0}^{\textrm{CT1}}
\equiv
A_{3}
\begin{bmatrix}
\cos\alpha_{3}\\
-
\sin\alpha_{3}
\end{bmatrix}
\end{split}
\end{equation}
with frequencies $\nu$ and $2\nu$. Under CT2, the limit cycle is mapped to the point,
\begin{equation}
\label{Eq: center point after CT2}
\begin{split}
\left( \theta^{\textrm{CT2}}, \tilde{p}_{\theta}^{\textrm{CT2}} \right)
=
Z_{0}^{\textrm{CT2}}
\equiv
A_{1}
\begin{bmatrix}
\cos\alpha_{1}\\
- \sin\alpha_{1}
\end{bmatrix}
\end{split}
\end{equation}
with frequencies $\nu$ and $2\nu$. Therefore, the upper stability boundary for the modulation strength $\lambda$ is determined by the critical value $\lambda_{c}$ at which these center points become nonzero stable fixed points of the effective generator.

To determine an explicit formula for $\lambda_{c}$, we derive the time-coarse grained equations of motion in the frames defined by CT1 and CT2. As a starting point, we write the original Hamiltonian and Dissipator in the transformed frames.

\begin{subequations}
\begin{equation}
\begin{split}
H^{\textrm{CT1}}
=&
-
\frac{9+4\Gamma^2}{36ml^{2}} \left( p_{\theta}^{\textrm{CT1}} \right)^{2}
-
\frac{ml^{2}\nu^{2}}{16} \left( A_{1}^{2}(1+4\Gamma^2-2\lambda) - 4A_{1}\lambda \theta^{\textrm{CT1}} + (9+4\Gamma^2) \left( \theta^{\textrm{CT1}} \right)^{2} \right)\\
&
+
\left(
\frac{\lambda}{9ml^{2}} \left( p_{\theta}^{\textrm{CT1}} \right)^{2}
+
\frac{ml^{2}\nu^{2}}{16}
\left( A_{1}^{2}(1-4\Gamma^2+4\lambda) - 2A_{1}(1+4\Gamma^2-2\lambda)\theta^{\textrm{CT1}} + 4\lambda \left( \theta^{\textrm{CT1}} \right)^{2} \right)
\right) \cos(\nu t)\\
&
-
\left(
\frac{\lambda}{18ml^{2}} \left( p_{\theta}^{\textrm{CT1}} \right)^{2}
-
\frac{ml^{2} \nu^{2}}{8} \left( \lambda A_{1}^{2} - A_{1}(1-2\lambda+4\Gamma^2) \theta^{\textrm{CT1}} + \lambda \left( \theta^{\textrm{CT1}} \right)^{2} \right)
\right) \cos(2\nu t)\\
&
+
\left(
\frac{9+4\Gamma^2}{36ml^{2}} \left( p_{\theta}^{\textrm{CT1}} \right)^{2}
+
\frac{ml^{2} \nu^{2}}{16} \left( 4A_{1}\lambda \cdot \theta^{\textrm{CT1}} - (9+4\Gamma^2) \left( \theta^{\textrm{CT1}} \right)^{2} \right)
\right) \cos(3\nu t)\\
&
-
\left(
\frac{\lambda}{18ml^{2}} \left( p_{\theta}^{\textrm{CT1}} \right)^{2}
-
\frac{ml^{2} \nu^{2} \lambda}{8} \left( \theta^{\textrm{CT1}} \right)^{2}
\right) \cos(4\nu t)\\
&
-
\frac{A_{1} \nu}{12} (1+4\Gamma^2-2\lambda) p_{\theta}^{\textrm{CT1}} \sin(\nu t)
+
\left(
\frac{\lambda \nu}{6} p_{\theta}^{\textrm{CT1}} \theta^{\textrm{CT1}}
-
\frac{(1+4\Gamma^2-2\lambda)\nu}{12} A_{1} p_{\theta}^{\textrm{CT1}}
\right) \sin(2\nu t)\\
&
-
\frac{\nu}{12} \left(
(9+4\Gamma^2) p_{\theta}^{\textrm{CT1}} \theta^{\textrm{CT1}}
-
2A_{1} \lambda p_{\theta}^{\textrm{CT1}}
\right) \sin(3\nu t)
+
\frac{\lambda \nu}{6} p_{\theta}^{\textrm{CT1}} \theta^{\textrm{CT1}} \sin(4\nu t)
\end{split}
\end{equation}

\begin{equation}
\begin{split}
D^{\textrm{CT1}}
=&
-
\frac{Q^{-1} \Gamma \nu}{16} \left(
A_{1}^{2} m^{2}l^{4} \nu^{2}
+
4 \left( p_{\theta}^{\textrm{CT1}} \right)^{2}
+
9m^{2}l^{4} \nu^{2} \left( \theta^{\textrm{CT1}} \right)^{2}
\right)
+
\frac{Q^{-1} \Gamma \nu^{3} m^{2}l^{4} A_{1}}{16} \left( A_{1} - 6\theta^{\textrm{CT1}} \right) \cos(\nu t)\\
&
-
\frac{Q^{-1} \Gamma\nu^{2}ml^{2}A_{1}}{4} p_{\theta}^{\textrm{CT1}} \sin(\nu t)
+
\frac{3Q^{-1} \Gamma\nu^{3}m^{2}l^{4}A_{1}}{8} \theta^{\textrm{CT1}} \cos(2\nu t)
+
\frac{Q^{-1} \Gamma\nu^{2}ml^{2}A_{1}}{4} p_{\theta}^{\textrm{CT1}} \sin(2\nu t)\\
&
-
\frac{Q^{-1} \Gamma \nu}{16} \left( 4\left( p_{\theta}^{\textrm{CT1}} \right)^{2} - 9m^{2}l^{4}\nu^{2} \left( \theta^{\textrm{CT1}} \right)^{2} \right) \cos(3\nu t)
+
\frac{3Q^{-1} \Gamma\nu^{2} ml^{2}}{4} p_{\theta}^{\textrm{CT1}} \theta^{\textrm{CT1}} \sin(3\nu t)
\end{split}
\end{equation}
\end{subequations}

\noindent\rule{\textwidth}{0.5pt}

\begin{subequations}
\begin{equation}
\begin{split}
H^{\textrm{CT2}}
=&
-
\frac{1+4\Gamma^2+2\lambda}{4ml^{2}} \left( p_{\theta}^{\textrm{CT2}} \right)^{2}
-
\frac{ml^{2}\nu^{2}}{16} \left( A_{3}^{2}(9+4\Gamma^2) - 4A_{3}\lambda \theta^{\textrm{CT2}} + (1+4\Gamma^2-2\lambda) \left( \theta^{\textrm{CT2}} \right)^{2} \right)\\
&
+
\left(
\frac{1+4\Gamma^2+4\lambda}{4ml^{2}} \left( p_{\theta}^{\textrm{CT2}} \right)^{2}
+
\frac{ml^{2}\nu^{2}}{16}
\left( 4A_{3}^{2}\lambda - 2A_{3}(9+4\Gamma^2-2\lambda)\theta^{\textrm{CT2}} - (1+4\Gamma^2-4\lambda) \left( \theta^{\textrm{CT2}} \right)^{2} \right)
\right) \cos(\nu t)\\
&
-
\left(
\frac{\lambda}{2ml^{2}} \left( p_{\theta}^{\textrm{CT2}} \right)^{2}
-
\frac{ml^{2} \nu^{2}}{8} \left( \lambda A_{3}^{2} - A_{3}(9-2\lambda+4\Gamma^2) \theta^{\textrm{CT2}} + \lambda \left( \theta^{\textrm{CT2}} \right)^{2} \right)
\right) \cos(2\nu t)\\
&
+
\frac{ml^{2} \nu^{2}}{16} A_{3} \left( A_{3}(9-4\Gamma^2) + 4\lambda \cdot \theta^{\textrm{CT2}} \right) \cos(3\nu t)
+
\frac{ml^{2} \nu^{2} \lambda}{8} A_{3}^{2} \cos(4\nu t)\\
&
+
\frac{\nu}{4} \left(
A_{3} (9+4\Gamma^2+2\lambda) p_{\theta}^{\textrm{CT2}} 
-
(1+4\Gamma^2) p_{\theta}^{\textrm{CT2}} \theta^{\textrm{CT2}}
\right) \sin(\nu t)\\
&
+
\frac{\nu}{4} \left(
2\lambda p_{\theta}^{\textrm{CT2}} \theta^{\textrm{CT2}}
-
(9+4\Gamma^2+2\lambda) A_{3} p_{\theta}^{\textrm{CT2}}
\right) \sin(2\nu t)
+
\frac{\lambda \nu}{2} A_{3} p_{\theta}^{\textrm{CT2}} \sin(3\nu t)
\end{split}
\end{equation}

\begin{equation}
\begin{split}
D^{\textrm{CT2}}
=&
-
\frac{Q^{-1} \Gamma \nu}{16} \left(
9A_{3}^{2} m^{2}l^{4} \nu^{2}
+
4 \left( p_{\theta}^{\textrm{CT2}} \right)^{2}
+
m^{2}l^{4} \nu^{2} \left( \theta^{\textrm{CT2}} \right)^{2}
\right)\\
&
+
\frac{Q^{-1} \Gamma \nu}{16} \left( 4\left( p_{\theta}^{\textrm{CT2}} \right)^{2} + m^{2}l^{4} \nu^{2} \left( 6A_{3} \theta^{\textrm{CT2}} - \left( \theta^{\textrm{CT2}} \right)^{2} \right) \right) \cos(\nu t)
+
\frac{3Q^{-1} \Gamma\nu^{3}m^{2}l^{4}}{8} A_{3} \theta^{\textrm{CT2}} \cos(2\nu t)\\
&
+
\frac{9Q^{-1} \Gamma\nu^{3}m^{2}l^{4}}{16} A_{3}^{2} \cos(3\nu t)
+
\frac{Q^{-1} \Gamma ml^{2} \nu^{2}}{4}
\left(
3A_{3} p_{\theta}^{\textrm{CT2}}
+
p_{\theta}^{\textrm{CT2}} \theta^{\textrm{CT2}}
\right) \sin(\nu t)
+
\frac{3Q^{-1} \Gamma \nu^{2} ml^{2}}{4} A_{3} p_{\theta}^{\textrm{CT2}} \sin(2\nu t)
\end{split}
\end{equation}
\end{subequations}

We write the effective generator to up different orders and derive the equations of motion of the time coarse-grained variables.
To first order, the equations of motion can be written as the following matrix equations, one for each frame:
\begin{equation}
\begin{split}
&
\partial_{t}
Z^{\textrm{CT1}}
=
-
M^{(1)}_{\textrm{CT1}}
\left(
Z^{\textrm{CT1}}
-
\left[ Z_{0}^{\textrm{CT1}} \right]^{(1)}
\right)\\
&
\partial_{t}
Z^{\textrm{CT2}}
=
-
M^{(1)}_{\textrm{CT2}}
\left(
Z^{\textrm{CT2}}
-
\left[ Z_{0}^{\textrm{CT2}} \right]^{(1)}
\right)
\end{split}
\end{equation}
where we define,
\begin{equation}
\begin{split}
&
Z^{\textrm{CT1}}
\equiv
\begin{bmatrix}
\theta^{\textrm{CT1}}\\
\tilde{p}_{\theta}^{\textrm{CT1}}
\end{bmatrix}
,\qquad
\left[ Z_{0}^{\textrm{CT1}} \right]^{(1)}
\equiv
\frac{2\lambda A_{1}}{(9+4\Gamma^2)^{2}+\left(6Q^{-1} \Gamma \right)^{2}}
\begin{bmatrix}
9+4\Gamma^2\\
- 6Q^{-1} \Gamma
\end{bmatrix}\\
&
Z^{\textrm{CT2}}
\equiv
\begin{bmatrix}
\theta^{\textrm{CT2}}\\
\tilde{p}_{\theta}^{\textrm{CT2}}
\end{bmatrix}
,\qquad
\left[ Z_{0}^{\textrm{CT2}} \right]^{(1)}
\equiv
\frac{2\lambda A_{3}}{(1+4\Gamma^2)^{2}-4\lambda^{2}+4Q^{-2}\Gamma^2}
\begin{bmatrix}
1+2\lambda+4\Gamma^2\\
- 2Q^{-1} \Gamma
\end{bmatrix}
\end{split}
\end{equation}

\begin{equation}
\begin{split}
M^{(1)}_{\textrm{CT1}}
\equiv
\begin{bmatrix}
\frac{1}{2} Q^{-1} \Gamma \nu,& \frac{9+4\Gamma^2}{12} \nu\\
-\frac{9+4\Gamma^2}{12} \nu ,& \frac{1}{2} Q^{-1} \Gamma\nu
\end{bmatrix}
,\qquad
M^{(2)}_{\textrm{CT2}}
\equiv
\begin{bmatrix}
\frac{1}{2} Q^{-1} \Gamma \nu,& \frac{1+2\lambda+4\Gamma^2}{4} \nu\\
-\frac{1-2\lambda+4\Gamma^2}{4} \nu ,& \frac{1}{2} Q^{-1} \Gamma \nu
\end{bmatrix}
\end{split}
\end{equation}

In order for $M^{(1)}_{\textrm{CT1}}$ and $M^{(2)}_{\textrm{CT2}}$ to be consistent with the limit cycle centers given in Eq.\ref{Eq: center point after CT1} and Eq.\ref{Eq: center point after CT2} has to satisfy the following self-consistency constraints,
\begin{equation}
\begin{split}
&
\frac{2\lambda A_{1}}{(9+4\Gamma^2)^{2}+\left(6Q^{-1} \Gamma\right)^{2}}
\begin{bmatrix}
9+4\Gamma^2\\
- 6Q^{-1} \Gamma
\end{bmatrix}
=
A_{3}
\begin{bmatrix}
\cos\alpha_{3}\\
-
\sin\alpha_{3}
\end{bmatrix}
\\
&
\frac{2\lambda A_{3}}{(1+4\Gamma^2)^{2}-4\lambda^{2}+4Q^{-2}\Gamma^{2}}
\begin{bmatrix}
1+2\lambda+4\Gamma^2\\
- 2Q^{-1} \Gamma
\end{bmatrix}
=
A_{1}
\begin{bmatrix}
\cos\alpha_{1}\\
- \sin\alpha_{1}
\end{bmatrix}
\end{split}
\end{equation}

Solutions to the above equations exist as long as the following self-consistency equation is satisfied:
\begin{equation}
\begin{split}
\frac{A_{3}^{2}}{A_{1}^{2}}
=
\frac{
4\lambda^{2}
}{
(9+4\Gamma^2)^{2}+\left(6Q^{-1} \Gamma\right)^{2}
}
=
\frac{
\left( (1+4\Gamma^2)^{2}-4\lambda^{2}+4Q^{-2}\Gamma^{2} \right)^{2}
}{
4\lambda^{2} \left( (1+2\lambda+4\Gamma^2)^{2}+4Q^{-2}\Gamma^{2} \right)
}
\end{split}
\end{equation}
which gives us the following standard result of Butikov et. al. \cite{Butikov2001} for $Q^{-1} =0$ that has been derived in the literature using what is effectively rotating wave approximations in the two rotating frames~\cite{Butikov2001}:
\begin{equation}
\begin{split}
\lambda_{c}^{(1)}\rvert_{Q^{-1}\rightarrow0}
=
\frac{\sqrt{117+232\Gamma^2+80\Gamma^{4}} - 9 - 4\Gamma^2}{4}
\end{split}
\end{equation}

For the dissipative case of $Q^{-1} \neq 0$, we can expand $\lambda_{c}$ to order $\mathcal{O}\left( \Gamma^4\right)$ and obtain the following approximate expression:
\begin{equation}
\label{Eq: lambda c order1}
\begin{split}
\lambda_{c}^{(1)}
\approx
0.454163
+
1.681051 \Gamma^2
+
0.859551 Q^{-2}\Gamma^{2}
-
0.404568 \Gamma^{4}
-
3.767032 Q^{-2} \Gamma^4
-
0.924046 Q^{-4} \Gamma^{4}
\end{split}
\end{equation}
In addition, it is straightforward to verify that the eigenvalues of $M^{(1)}_{\textrm{CT1}}$ and $M^{(2)}_{\textrm{CT2}}$ all have positive real parts under our assumption that $\Gamma\ll 1$ when $\lambda = \lambda_{c} \sim 0.454$, which ensures that the limit cycle is stable.

Going beyond the first-order dynamics, one obtains corrections to this standard result. For instance, at the third order one has the following equation at the limit $Q^{-1}, \Gamma \rightarrow 0$:
\begin{equation}
\begin{split}
\frac{A_{3}^{2}}{A_{1}^{2}}
=
\frac{
4\lambda^{2} \left( 252 + \lambda(20\lambda-9) \right)^{2}
}{
81\left( 252 + (5-4\lambda) \lambda^{2} \right)^{2}
}
=
\frac{
36
\left(
7
-
\lambda\left( 12 + 3\lambda - 7\lambda^{2} \right)
\right)^{2}
}{
\lambda^{2} \left(
138
+
159 \lambda
-
10 \lambda^{2}
\right)
}
\end{split}
\end{equation}
which gives us $\lambda_{c}^{(3)}\rvert_{Q^{-1}\rightarrow0} \approx 0.454203$.
Up to order $\mathcal{O}\left( \Gamma^4 \right)$, we have
\begin{equation}
\label{Eq: lambda c order3}
\begin{split}
\lambda_{c}^{(3)}
\approx
0.454203
+
1.681404 \Gamma^2
+
0.883543 Q^{-2} \Gamma^{2}
-
0.409313 \Gamma^4
-
3.049322 Q^{-2} \Gamma^4 
-
0.743824 Q^{-4} \Gamma^{4}
\end{split}
\end{equation}

Our method enables us to systematically calculate higher-order contributions. To complete the analysis, we also examine the fifth-order contribution.
\begin{equation}
\label{Eq: lambda c order5}
\begin{split}
\lambda_{c}^{(5)}
\approx
0.454249
+
1.681937 \Gamma^2
+
0.817923 Q^{-2} \Gamma^{2}
-
0.412537 \Gamma^{4}
-
3.437559 Q^{-2} \Gamma^4 
-
0.983173 Q^{-4} \Gamma^{4}
\end{split}
\end{equation}
In this particular case, the convergence of the effective generator is rather slow, and the coefficients in Eq.(\ref{Eq: lambda c order1},\ref{Eq: lambda c order3},\ref{Eq: lambda c order5}) converge roughly at the same rate as the alternating harmonic series, which makes the closed-form expression at high orders a crucial result.

\subsubsection{The parametrically-modulated Kapitza pendulum}

As we discuss in the main text, our new formulation of the time-coarse graining framework in terms of generalized cumulants of Lie derivatives makes it straightforward, among other features, to treat Hamiltonian and dissipators with the most general time-dependence. As a demonstration, we consider another variation of the classical Kapitza pendulum described by the following Hamiltonian and dissipator:
\begin{subequations}
\begin{equation}
\begin{split}
H(t)
=&
\frac{p_{\theta}^{2}}{2m l(t)^{2}}
+
m l(t)
\big(  g - \lambda l(t) \nu^{2} \cos( \nu t ) \big)
\cos\theta
\end{split}
\end{equation}
\begin{equation}
\begin{split}
D(t)
=
-
\frac{1}{2}
Q^{-1} \sqrt{\frac{g}{l(0)}} p_{\theta}^{2}
\equiv
-
\frac{1}{2}
Q^{-1} \Gamma \nu p_{\theta}^{2}
\end{split}
\end{equation}   
\end{subequations}
where now the length of the pendulum is modulated in time.

During the beginning of the modulation time ($t\sim t_0)$, we can write,
\begin{equation}
\begin{split}
l(t)
=
l(t_{0})\big(1 + \Delta(t) \big)
\end{split}
\end{equation}
where $\Delta(t)$ is a non-adiabatic modulation of the length,
\begin{equation}
    \Delta(t) \equiv \alpha_{1}(t_{0}) (t-t_{0}) + \alpha_{2}(t_{0}) (t-t_{0})^{2}   
\end{equation}

We will assume that the modulation of the length is weak, allowing us to expand the generator to linear order in $\alpha_{1,2}$,
\begin{subequations}
\begin{equation}
\begin{split}
H(t)
\approx&
\frac{p_{\theta}^{2}}{2m l^{2}}
-
\frac{p_{\theta}^{2}}{m l^{2}} \Delta
+
m g l(t_{0})
\left(
1 + \Delta - \Gamma(t)^{-2}\lambda(1+2\Delta) \cos( \nu t )
\right)
\cos\theta
\end{split}
\end{equation}
\begin{equation}
    D(t)
=
-
\frac{1}{2}
Q^{-1} \Gamma(t) \nu p_{\theta}^{2}   
\end{equation}    
\end{subequations}
Note that due to the length modulation, the natural frequency $\omega_0$ and $\Gamma$ now change in time as well.

We will assume that the length modulation is much slower than the vertical driving period $\nu^{-1}$, guaranteeing a separation of time-scales. In order to filter out the vertical modulation while capturing the slow non-adiabatic changes, we choose our time coarse-graining scale such that $\nu^{-1} \ll \tau \ll 1/\alpha_{1}(t_{0}), 1/\sqrt{\alpha_{2}(t_{0})}$ for some slowly varying $l(t_{0})$, $\alpha_{1}(t_{0})$ and $\alpha_{2}(t_{0})$. This allows us to write,
\begin{equation}
\begin{split}
\overline{\Delta}(t_{0})
\approx
\alpha_{2}(t_{0}) \tau^{2}
,\qquad
\overline{\dot{\Delta}}(t_{0})
\approx
\alpha_{1}(t_{0})
\end{split}
\end{equation}

Using our double-bracket formulation, for any function $z(\theta, p_{\theta})$ in phase-space we have,
\begin{equation}
\begin{split}
\partial_{t} z
=&
\left\{ z, H(t) \right\}
+
\left\{\left\{ z, D(t) \right\}\right\}
\equiv
\left(
\frac{\partial z }{\partial \theta} \frac{\partial H(t) }{\partial p_{\theta}}
-
\frac{\partial z }{\partial p_{\theta}} \frac{\partial H(t) }{\partial \theta}
\right)
+
\frac{\partial z }{\partial p_{\theta}} \frac{\partial D(t) }{\partial p_{\theta}}
\end{split}
\end{equation}

We calculate the effective generators $\mathcal{U}^{(k)}$ up to the third order, and their action on the (dimensionless) phase space coordinates
$\big( \theta, \tilde{p}_{\theta} \big)
\equiv
\big( \theta, p/l(t)^{2}m \big)$. Remarkably, for this problem the corrections split simply into double-bracket dissipators and effective Hamiltonian, although this is generally not the case. These effective Hamiltonian and dissipator terms can be determined uniquely at each order if one assumes that the time evolution on the phase space is given by
\begin{equation}
\begin{split}
\partial_{t} z
=&
\left\{ z, H^{(1)}_{\textrm{eff}}(t)+H^{(2)}_{\textrm{eff}}(t)+H^{(3)}_{\textrm{eff}}(t)+\cdots \right\}
+
\left\{\left\{ z, D^{(1)}_{\textrm{eff}}(t)+D^{(2)}_{\textrm{eff}}(t)+D^{(3)}_{\textrm{eff}}(t)+\cdots \right\}\right\}\\
\equiv&
\left(
\frac{\partial z }{\partial \theta} \frac{\partial H(t) }{\partial p_{\theta}}
-
\frac{\partial z }{\partial p_{\theta}} \frac{\partial H(t) }{\partial \theta}
\right)
+
\frac{\partial z }{\partial p_{\theta}} \frac{\partial D(t) }{\partial p_{\theta}}
\end{split}
\end{equation}
for $z\in\{\theta, p_{\theta}\}$
where
\begin{equation}
\begin{split}
\left\{
f,g
\right\}
\equiv
\frac{\partial f }{\partial \theta} \frac{\partial g }{\partial p_{\theta}}
-
\frac{\partial f }{\partial p_{\theta}} \frac{\partial g }{\partial \theta}
,\qquad
\left\{\left\{
f,g
\right\}\right\}
\equiv
\frac{\partial f }{\partial p_{\theta}} \frac{\partial g }{\partial p_{\theta}}
.
\end{split}
\end{equation}

At first-order, the effective dynamics are given simply by the averaged generators,
\begin{equation}
\begin{split}
&
\mathcal{U}^{(1)} \theta
=
\Big(
1
-
2 \overline{\Delta(t)}
\Big)
\tilde{p}_{\theta}\\
&
\mathcal{U}^{(1)} \tilde{p}_{\theta}
=
-
Q^{-1} \Gamma \nu \tilde{p}_{\theta}
+
\Gamma^2(t) \nu^{2} \left( 1 + \overline{\Delta(t)} \right)
\sin\theta
\end{split}
\end{equation}

\begin{equation}
\begin{split}
H^{(1)}_{\textrm{eff}}(t)
=
\overline{H}(t)
,\qquad
D^{(1)}_{\textrm{eff}}(t)
=
\overline{D}(t)
\end{split}
\end{equation}

At the second order, we obtain new corrections due to the time-dependent nature of the length modulation.
\begin{equation}
\begin{split}
&
\mathcal{U}^{(2)} \theta
=
\Gamma^2(t) \nu^{2} \tau^{2} \overline{\dot{\Delta}(t)} \sin\theta\\
&
\mathcal{U}^{(2)} \tilde{p}_{\theta}
=
-
\Gamma^2(t) \nu^{2}\tau^{2} \overline{\dot{\Delta}(t)}
\big(
2\left(\cos\theta\right) \tilde{p}_{\theta}
+
Q^{-1} \Gamma \nu \sin\theta
\big)
\end{split}
\end{equation}
\begin{equation}
\begin{split}
&
H^{(2)}_{\textrm{eff}}(t)
=
\Gamma^2 \tau^{2} \nu^{2} \overline{\dot{\Delta}(t)}
\left[
\left( \sin\theta \right) p_{\theta}
-
Q^{-1} \Gamma \nu m l(t)^{2} \cos\theta
\right]\\
&
D^{(2)}_{\textrm{eff}}(t)
=
-
\frac{1}{2} \Gamma^2 \tau^{2} \nu^{2}
\overline{\dot{\Delta}(t)} \cos\theta
p_{\theta}^{2}
\end{split}
\end{equation}

At third-order we see additional non-adiabatic effects, arising from higher time-derivative of the modulation,
\begin{equation}
\begin{split}
&
\mathcal{U}^{(3)} \theta
=
-
\Gamma^2(t) \alpha_{2} \nu^{2} \tau^{4}
\big(
2\left(\cos\theta\right) \tilde{p}_{\theta}
+
Q^{-1} \Gamma \nu \left(\sin\theta\right)
\big)\\
&
\mathcal{U}^{(3)} \tilde{p}_{\theta}
=
2\alpha_{2} \Gamma^2(t) \nu^{2} \tau^{4}
\left(
Q^{-1} \Gamma \nu \left(\cos\theta\right) \tilde{p}_{\theta}
+
\left(\sin\theta\right) \tilde{p}_{\theta}^{2}
\right)
+
\alpha_{2} \Gamma^2(t) Q^{-2} \Gamma^{2} \nu^{4} \tau^{4} \sin\theta\\
&\qquad\qquad
-
\left(
\frac{\lambda^{2}\nu^{2}}{4}
+
\frac{1}{2} \left(
\lambda^{2} \nu^{2} \overline{\Delta(t)}
-
\alpha_{2} \left( 4\lambda^{2} + \Gamma^2(t)^{2} \nu^{4} \tau^{4} \right)
\right)
\right) \sin(2\theta)
\end{split}
\end{equation}

\begin{equation}
\begin{split}
&
H^{(3)}_{\textrm{eff}}(t)
=
-
\Gamma^2 \alpha_{2} \nu^{2} \tau^{4}
\left(
\left( \cos\theta \right) \frac{p_{\theta}^{2}}{m l(t)^{2}}
+
Q^{-1} \Gamma \nu \left( \sin\theta \right) p_{\theta}
\right)\\
&\qquad\qquad
+
m l(t)^{2}
\left(
\alpha_{2} Q^{-2}\Gamma^{2} \Gamma^2 \nu^{4}\tau^{4} \cos\theta
-
\left(
\frac{\lambda^{2} \nu^{2}}{8}
+
\frac{\overline{\Delta(t)}\lambda^{2}\nu^{2}}{4}
-
\alpha_{2}\left( \lambda^{2} + \frac{\Gamma^4 \nu^{4} \tau^{4}}{4} \right)
\right) \cos(2\theta)
\right)\\
&
D^{(3)}_{\textrm{eff}}(t)
=
-
\frac{1}{2}
\left[
-
\alpha_{2} \Gamma^2 \nu^{2} \tau^{4}
\left(
Q^{-1} \Gamma \nu \left( \cos\theta \right) p_{\theta}^{2}
+
\left( \sin\theta \right) \frac{2p_{\theta}^{3}}{m l(t)^{2}}
\right)
\right]
\end{split}
\end{equation}

Based on the explicit expressions above, we note the following features of the effective TCG dynamics: The first-order TCG dynamics is generated by the time-averaged Hamiltonian and dissipator. Beyond the first order, the corrections to $\mathcal{U}^{(1)}(t)$ cannot be fully encapsulated in the form of corrections to the effective Hamiltonian in general. The remaining terms in $\mathcal{U}(t)\theta$ and $\mathcal{U}(t)p_{\theta}$ can be obtained from additional dissipator terms, which can be considered as due to non-adiabaticity since they vanish in the $\alpha_{1},\alpha_{2} \rightarrow 0$ limit. The TCG corrections are generally dependent on the coarse-graining time scale $\tau$, although some terms are independent of $\tau$ and may therefore survive the $\tau \rightarrow 0$ limit if one keeps the assumption $\nu^{-1} \ll \tau$. These corrections can be considered as due to ultra-fast virtual processes.

\begin{figure*}[h!]
    \centering
    \includegraphics[width=\linewidth]{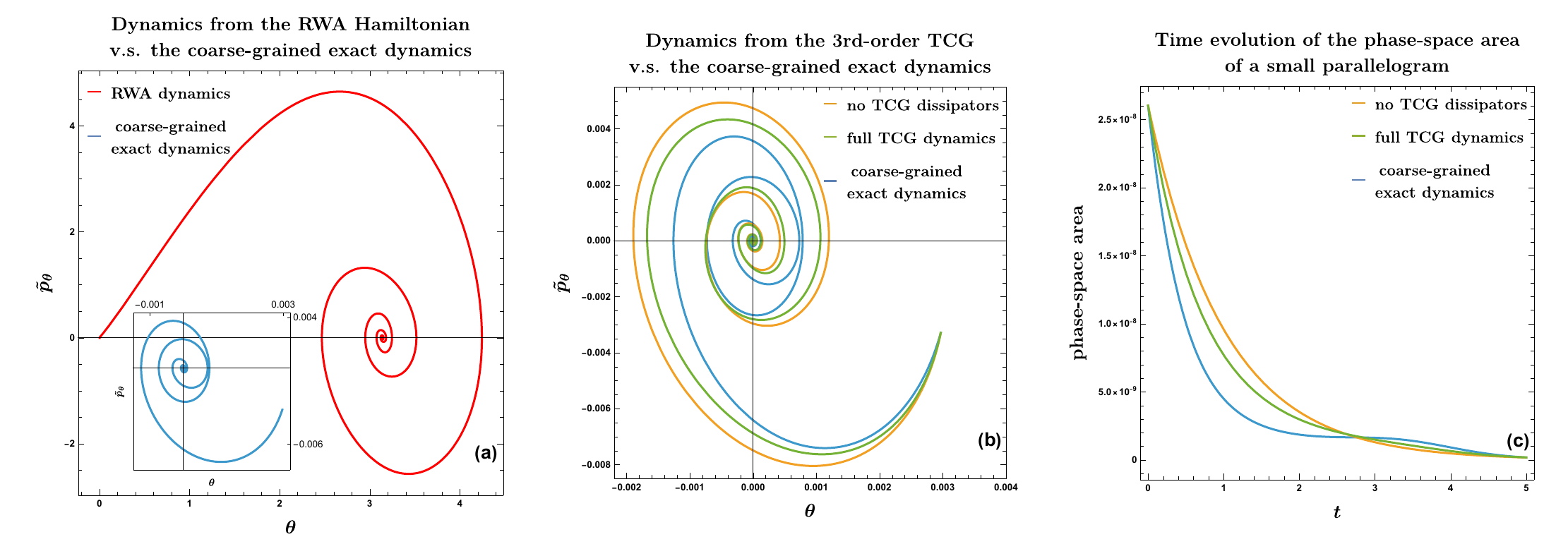}
    \caption{\textbf{(a)} The time-coarse grained exact dynamics is compared with that obtained by removing the fast-oscillating drive in the Hamiltonian. The secular behaviors of the two are qualitatively different as the RWA dynamics cannot capture the dynamical stabilization of fixed points. \textbf{(b)} Dynamical stabilization is captured by third-order TCG. In addition to the effective Hamiltonian, corrections in the form of effective dissipators arise during TCG, which improves the accuracy of the predicted secular dynamics. \textbf{(c)} The effective dissipators give rise to additional modulation of the phase-space area (volume) on top of that due to the built-in dissipation. This additional variation in the phase-space area (volume) is due to the coarse-graining procedure and not any dissipative mechanism at the fundamental level.}
    \label{fig:Kapitza_real_time}
\end{figure*}

To demonstrate the validity of our description, we simulate the time evolution with,
\begin{equation}
\begin{split}
l(t)
=
l_{0}\left(
1
+
A\sin(t/T)
\right)
\end{split}
\end{equation}
and the following model parameters:
\[
\Gamma^2 = 0.02,
\lambda = 0.3,
\nu = 20,
Q^{-1} \Gamma = 0.05,
A = 0.2,
T = 5/2\pi
.
\]
As an arbitrary initial condition for our numerical simulation,  we choose,
\[
\theta(-2) = 0.01,\quad \tilde{p}_{\theta}(-2) = 0
\]

The fast dynamics can be effectively filtered out by choosing a coarse-graining time-scale $\tau = 0.4$ Fig.~\ref{fig:Kapitza_real_time} shows how the RWA dynamics, obtained by removing the rapidly-oscillating terms in the Hamiltonian, fails to capture the dynamical stabilization of the point $(\theta, \tilde{p}_{\theta}) = (0,0)$. Starting at the third order, time-coarse graining gives rise to equations of motion capable of capturing the long-time dynamics and dynamical stabilization. Additionally, the dissipator terms produced by time-coarse graining make the simulated secular dynamics more accurate by accounting for the additional variations in the phase-space area (volume) due to time-coarse graining.

\subsection{Worked Example II -- the parametrically driven oscillator}

The parametric oscillator serves as a paradigmatic system for exploring time-dependent Hamiltonian dynamics, stability, and energy transfer mechanisms in classical and quantum physics. Unlike conventional driven oscillators, where external forces act additively, parametric oscillators are modulated through their internal parameters. For example, we consider a harmonic oscillator with natural frequency $\omega_0$ which  is modulated in time. 
\begin{equation}
    	H(t; \epsilon) = \frac{p^2}{2m} + \frac{1}{2} m\omega^2_0 q^2(1 + \epsilon \sin\Omega t).
\end{equation}
where $\varepsilon$ is the modulation strength. This modulation leads to rich phenomena such as exponential amplification, threshold behavior, and parametric resonance. These features make the system a fertile ground for studying fundamental questions in dynamical stability, mode coupling, and energy localization. In quantum settings, parametric oscillators underpin technologies ranging from squeezed light generation to quantum-limited amplification.

It is often convenient to write the Hamiltonian in terms of the action-angle variables,
\begin{equation}
    	q = \sqrt{\frac{2J}{m\omega_0}} \cos\theta, \quad p = - \sqrt{2m\omega_0 J} \sin\theta
\end{equation}
Plugging in, the Hamiltonian transforms into,
\begin{equation}
    	H(t; \epsilon) = \omega_0J + \epsilon \omega_0\sin \Omega t \cdot J \cos^2\theta
\end{equation}

To facilitate the application of our averaging procedure, we transform to a frame in which the coordinates vary slowly. In this case, since the oscillator is nearly harmonic, the motion is expected to be dominated by a rapid rotation at the natural frequency $\omega_0$, suggesting that a co-rotating frame will capture the slowly-varying dynamics more effectively. We can do that by transforming to a rotating-frame,
\begin{equation}
    R_{\omega_0} = \exp (-\mathcal{L}_{\omega_0 J} t)
\end{equation}

In the new variables $\tilde{z}$,
\begin{equation}
    \frac{\rm{d} \tilde{z}}{\rm{d}t} = -\mathcal{L}_{\omega_0 J} \tilde{z} + \mathcal{L}_{R_{\omega_0} H} \tilde{z} = \mathcal{L}_{H_1} \tilde{z}.
\end{equation}
which defines the new transformed Hamiltonian,
\begin{equation}
    H_1(t;\epsilon) = \epsilon \omega_0 \sin \Omega t \cdot J \cos^2(\theta + \omega_0 t)
\end{equation}
Note that the form of $H_1 = \varphi(q_l, p_l) \sin(\Omega t)$ is quite generic, and this type of driving can describe various fundamental phenomena, such as parametric resonances and ponderomotive forces, which are important in both classical and quantum systems. In many instances, the simple parametric oscillator emerges as a limiting case when the dynamics of the drive can be neglected. 

We can now easily write the Hamiltonian as a sum of terms with a simple external time-dependence, by expanding it in a Fourier series,
\begin{align*}
    	H(t; \epsilon) &= \frac{\epsilon}{2}\omega_0 J \cos \Omega t &\qquad (\omega=\Omega)\\
	&+ \frac{\epsilon}{4} \omega_0 J \left(\cos 2\theta \cdot \cos\Sigma t + \sin2\theta \cdot \sin \Sigma t  \right)  &\qquad (\omega=\Sigma) \\
	&+ \frac{\epsilon}{4} \omega_0 J \left(\cos 2\theta \cdot \cos\Delta t + \sin2\theta \cdot \sin \Delta t  \right)  &\qquad (\omega=\Delta)
\end{align*}
where we defined the sum and difference frequencies,
\begin{align}
	\Sigma &\equiv \Omega + 2\omega_0 \\
	\Delta &\equiv \Omega - 2\omega_0 
\end{align}

It is useful to write the Hamiltonian in terms of a complex Fourier series $H(t;\epsilon) = \sum_{\omega \in \nu} h_\omega(\epsilon) e^{-i\omega t}$, making the averaging easier. Separating the Hamiltonian into its frequency components, we have,
\begin{align*}
	&h_{\Omega} = \frac{\epsilon}{4} \omega_0 J = h_{-\Omega} \\
	&h_\Delta = \frac{\epsilon}{8} \omega_0 J e^{-2i\theta} = h_{-\Delta}^* \\
	&h_{\Sigma} = \frac{\epsilon}{8} \omega_0 J e^{-2i\theta} = h_{-\Sigma}^*
\end{align*}

\subsubsection{First-order corrections}
To proceed with the calculation, we must define which frequencies are considered \emph{slow} and which are \emph{fast}. We assume a near-resonance condition, such that \(\Delta \ll \Sigma, \Omega\). This results in a clear separation of timescales, allowing us to neglect high-frequency terms within our order of approximation without sacrificing accuracy.

The first-order corrections are straightforward, consisting solely of single-letter Lie words:
\begin{equation}
    \mathcal{U}_1 = \sum_{\omega \in \{\pm\Sigma,\pm\Delta\}} U_\omega \mathcal{L}_{h_{w}} = \sum_{\omega \in \{\pm\Sigma,\pm\Delta\}}  \overline{e^{-i\omega t}}\mathcal{L}_{h_{w}}
\end{equation}
The first-order cumulants $U_\omega$ are simply the coarse grained time-dependent functions. The first-order correction to the dynamics is thus simply the time-averaged Hamiltonian, with each frequency component weighted by the spectral filter \(\overline{e^{-i\omega t}}\). By construction, these first-order contributions reduce to a sum of Lie derivatives and are therefore manifestly Hamiltonian.

To maintain the accuracy of the approximation, we retain all frequency components bigger than $O(\epsilon)$. If the frequencies \(\Sigma\) and \(\Omega\) are significantly larger than the difference frequency \(\Delta\), as is the case under near-resonance, they can be safely neglected. Consequently, only the difference-frequency components contribute:
\begin{equation}
    K_1 = \frac{1}{4} \epsilon \omega_0 \cdot J \left( \cos 2\theta \cos \Delta t - \sin 2\theta \sin \Delta t \right).
\end{equation}
Depending on the system's parameters, this term can lead to either instability or amplification. However, this simplification breaks down under large detuning, where the full Hamiltonian must be retained to preserve reversibility. In such cases, the resulting Hamiltonian remains rapidly time-dependent and essentially equivalent to the original, up to attenuation by the spectral window. This indicates that the chosen filter (a low-pass centered around \(\omega_0\)) is no longer suitable for describing the system's dynamics, and a different filtering strategy must be employed.

The same first-order correction can also be obtained using Lie perturbation theory (LPT). In LPT, we derive two Hamiltonians -- $W$ which generates the simplifying canonical transformation, and $K$ (also known as the Kamiltonian), which generates the slow-time evolution of the system. In LPT, one has the freedom to assign which frequencies appear in \(K\), provided that the generator $W$ and the Kamiltonian \(K\) are spectrally well-separated. This freedom is analogous to the choice of filter in time-coarse-graining (TCG); however, the methods differ in how they handle the discarded frequencies. TCG removes high-frequency terms entirely, while LPT retains them in the generator of the symplectic transformation.

In our example, we retain the difference-frequency components in the Kamiltonian \(K\), and the high-frequency harmonics are retained within the generator:
\begin{equation}
    w_1 = \epsilon \omega_0 J \left( \frac{\cos(\Omega t)}{2\Omega} + \frac{1}{4} \frac{\cos(\Sigma t + 2\theta)}{\Sigma} \right).
\end{equation}
This highlights a key distinction between TCG and LPT: LPT never discards information. Harmonics excluded from the transformed Hamiltonian reappear in the inverse transformation and manifest in the lab-frame observables. To first order, the lab-frame variables are given by $z = \overline{z} + \{w_1(\overline{z}, t), \overline{z}\}$, ensuring that no dynamical content is lost in the transformation.

\subsubsection{Higher-order corrections}

\begin{figure*}
    \centering
    \includegraphics[width=\linewidth]{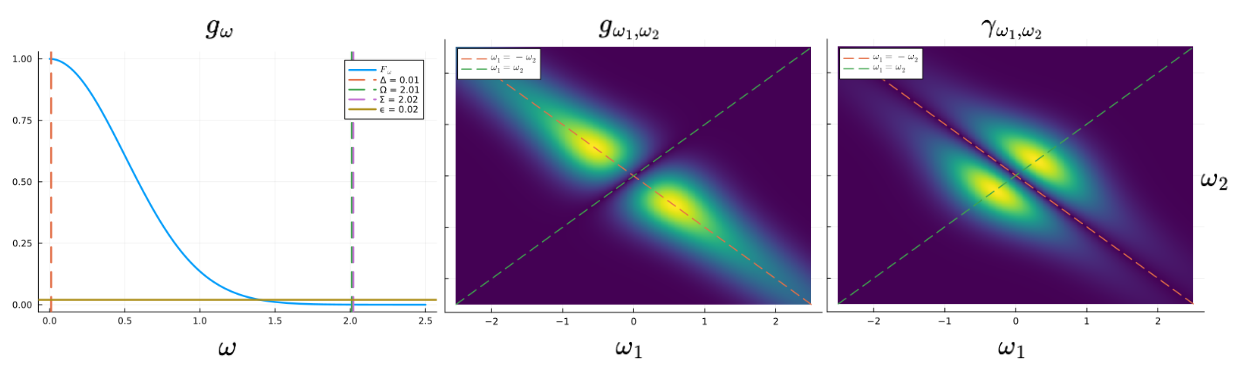}
    \caption{\textbf{(Left)} First-order couplings -- the width of the window is chosen such that negligible contributions are small. For near resonance operation, only the difference frequency $\Delta$ is significant. \textbf{(Middle)} Magnitude of second-order non-Hamiltonian couplings $\gamma_{\omega_1, \omega_2}$ for a Gaussian window function. \textbf{(Right)} Magnitude of Hamiltonian contributions $g_{\omega_1, \omega_2}$. By symmetry, the red $(\omega, -\omega)$ line is purely Hamiltonian, while the blue $(\omega, \omega)$ line is purely non-Hamiltonian. The couplings depend only on the problem’s spectrum and the window function, independent of physical details.}
    \label{fig:tcg_couplings}
\end{figure*}

In second-order, the generator is more complicated, now comprised of two-letter "words".
\begin{subequations}
\begin{equation}
    \mathcal{U}_2 = \sum_{\omega_1,\omega_2 \in \{\pm\Delta, \pm\Sigma, \pm\Omega\}} U_{\omega_1 \omega_2} \mathcal{L}_{h_{\omega_1}} \mathcal{L}_{h_{\omega_2}}
\end{equation}
\begin{equation}
    U_{\omega_1 \omega_2} = \frac{i}{\omega_2} \left( \overline{e^{-i(\omega_1 + \omega_2)t}} - \overline{e^{-i\omega_1 t}} \, \overline{e^{-i\omega_2 t}}  \right)
\end{equation}
\end{subequations}

Calculating $U_{\omega_1 \omega_2}$, we observe that the only contributions are from words whose frequencies are small, in the form of the conjugate words $w = (\omega, -\omega)$ and the symmetric words $w = (\Delta, \Delta)$.

As illustrated in Fig.~\ref{fig:tcg_couplings}, the conjugate words contribute only a Hamiltonian term, obtained by summing over their Lie-contributions,

\begin{equation}
    K_2 = \left(\frac{1}{2\Delta} + \frac{1}{2\Sigma} \right) \epsilon^2 \omega_0^2 J \approx \frac{1}{16} \epsilon^2 \omega_0^2 \Delta \cdot J, \quad \Delta \sim 0
\end{equation}
This part of the second-order contribution is a detuning-dependent frequency shift, which vanishes on resonance. Again, this is identical to the correction obtained by LPT. 

However, unlike LPT, TCG produces additional, non-Hamiltonian terms, arising from the symmetric word $w = (\Delta, \Delta)$. 
\begin{equation}
 	\mathcal{D}_{\Delta \Delta} = \frac{\epsilon^2 \omega_0^2}{32} e^{-4i\theta} \left[  \left(\partial_\theta^2 - 4J^2 \partial_J^2 \right) + i\left( 4J \partial_{\theta J} - 2\partial_\theta \right) \right]   
\end{equation}

It is useful to divide into real and imaginary parts,
\begin{align}
	D_{\Delta\Delta}^{re} = \frac{\epsilon^2\omega_0^2}{32} \left[ \cos 4\theta \cdot\mathcal{I} + \sin 4\theta \cdot \mathcal{Q} \right] \\
	D_{\Delta \Delta}^{im} = \frac{\epsilon^2\omega_0^2}{32} \left[ \cos 4\theta \cdot\mathcal{Q} - \sin 4\theta \cdot \mathcal{I} \right]
\end{align}
where we define,
\begin{align}
	&\mathcal{I} = \partial_\theta^2 - 4J^2 \partial_J^2 \\
	&\mathcal{Q} = 4J \partial_{\theta J} - 2\partial_\theta
\end{align}

\begin{figure}
     \centering
    \includegraphics[width=0.49\linewidth]{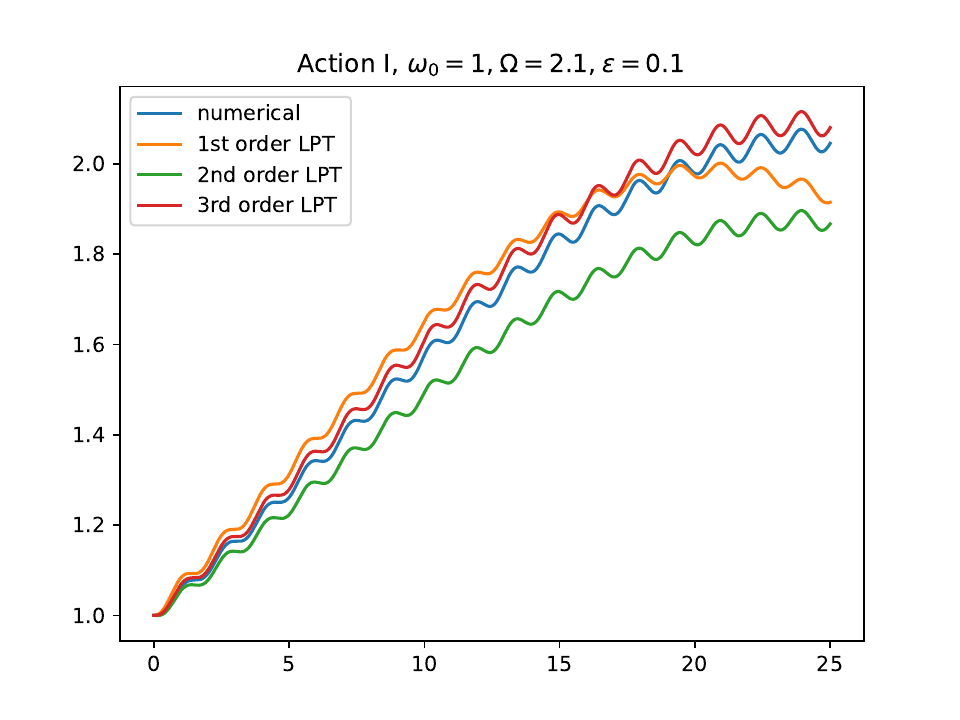}
    \includegraphics[width=0.49\linewidth]{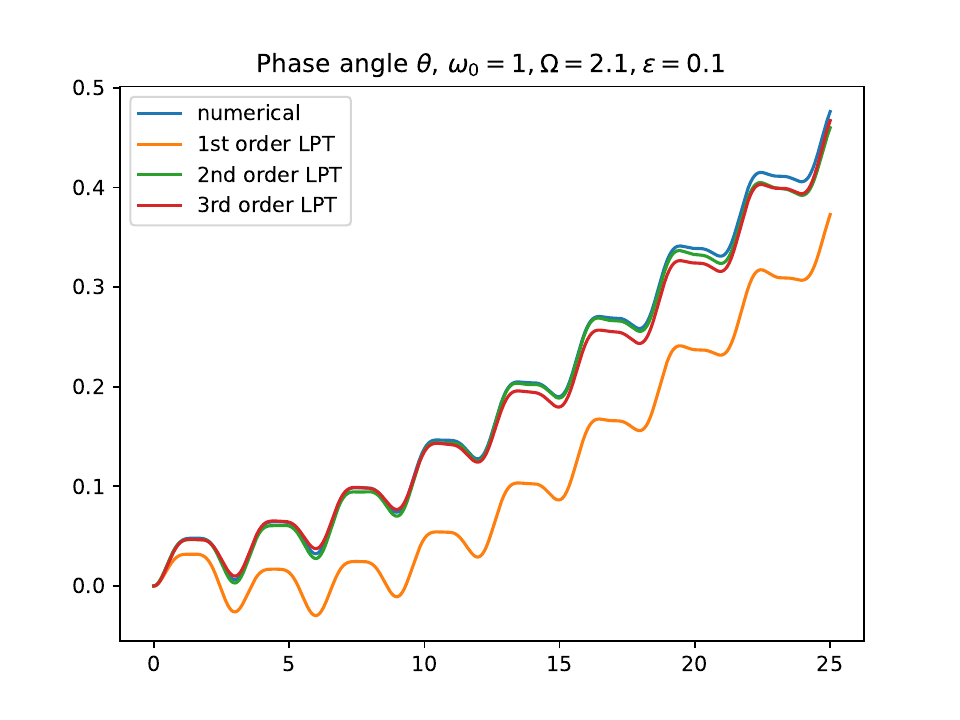}
     \caption{The action integral and phase angel evaluated at three orders of approximation using the Lie perturbation theory away from resonance.}
     \label{fig:LPTj}
 \end{figure}






 \begin{figure}
     \centering
 \includegraphics[width=0.49\linewidth]{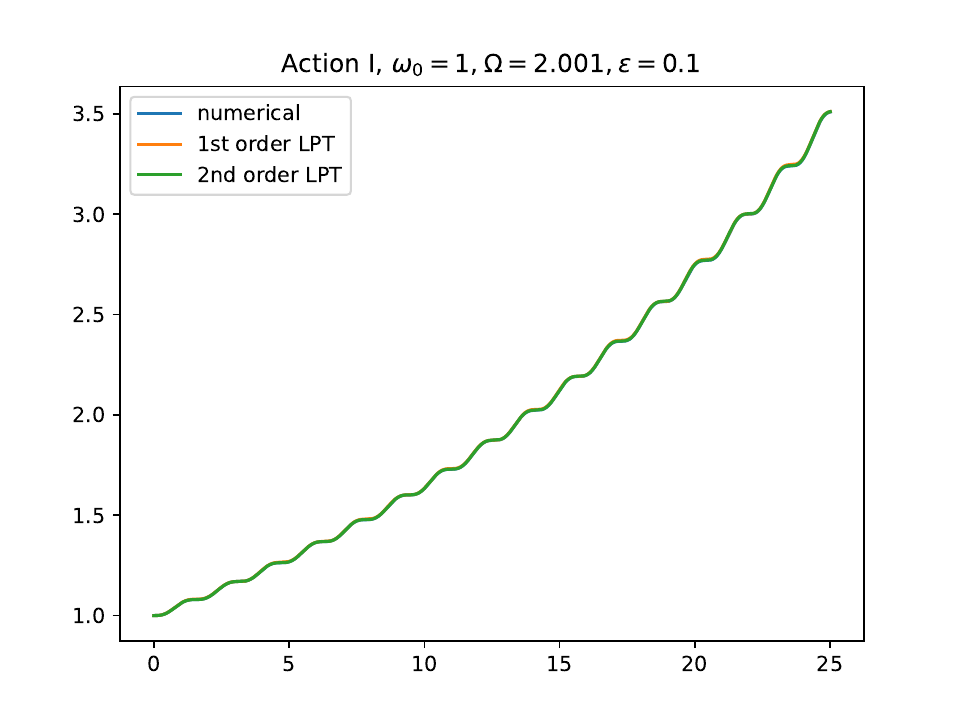}
 \includegraphics[width=0.49\linewidth]{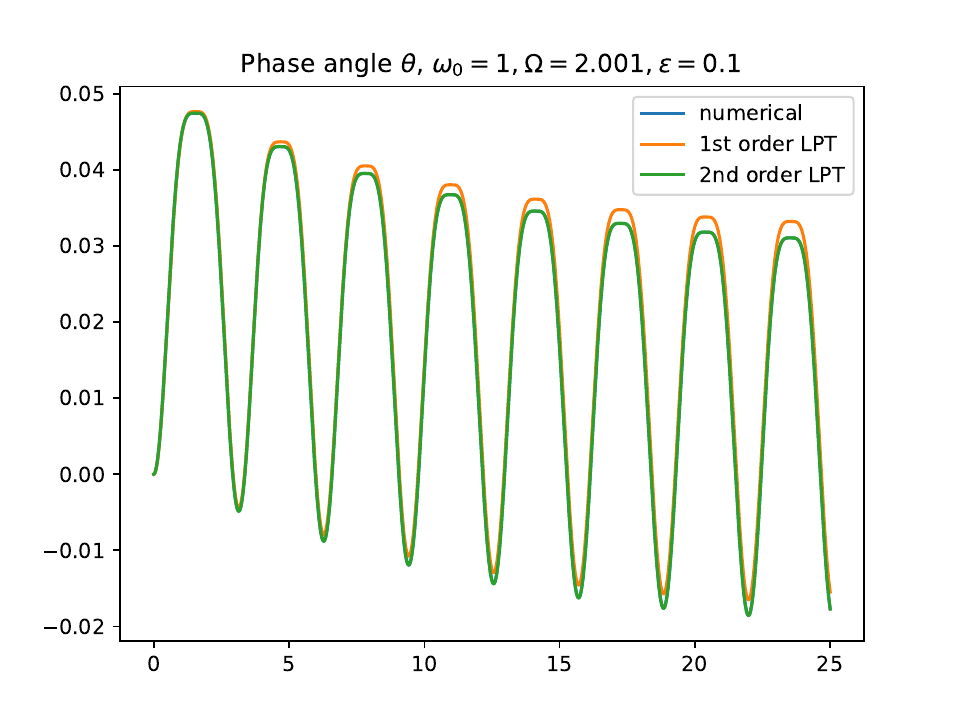}
     \caption{The action integral and phase angle evaluated at two orders of approximation using the Lie perturbation theory next to resonance. For the action, the three lines practically overlap.}
     \label{fig:LPTjres}
 \end{figure}



 \begin{figure}
     \centering
 \includegraphics[width=0.49\linewidth]{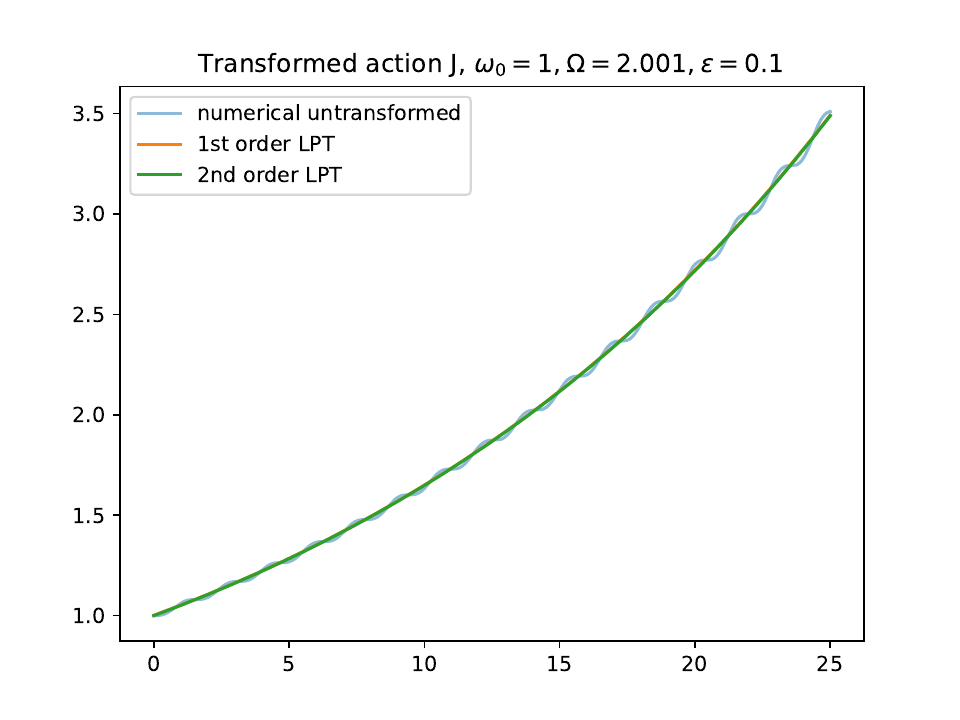}
  \includegraphics[width=0.49\linewidth]{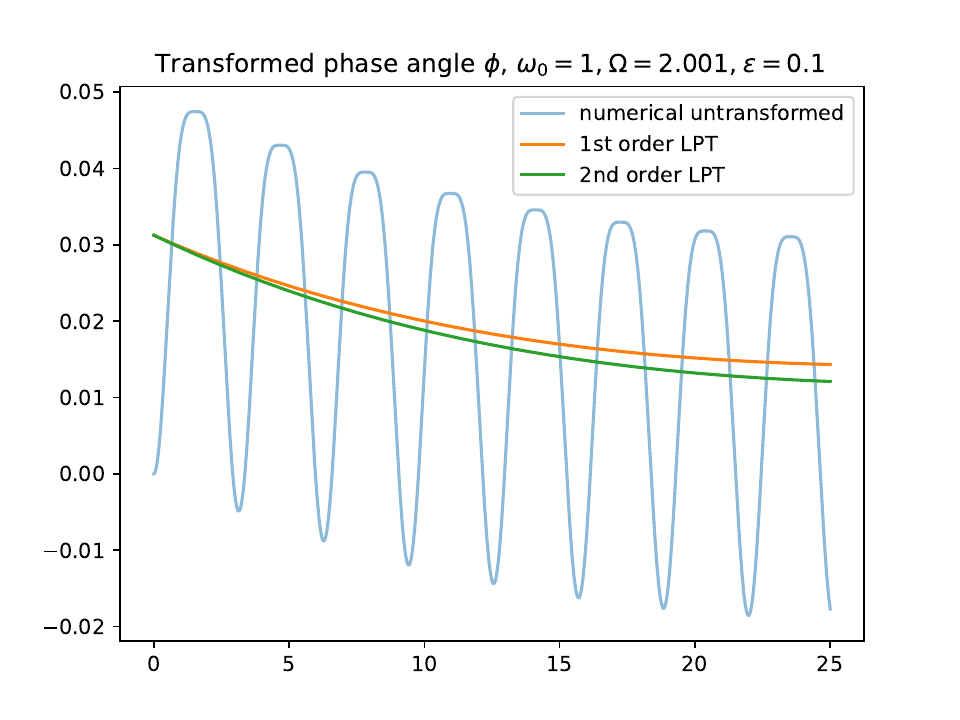}

     \caption{The transformed action integral J and phase angle $\phi$ close to resonance, evaluated at two orders of approximation using the Lie perturbation theory, compared to the untransformed numerical solution.}
     \label{fig:LPTIres}
\end{figure}


Since it is difficult to interpret the effects of the non-Hamiltonian terms from the generator, we look at the equations of motion of $(J, \theta)$,
\begin{subequations}
\begin{equation}
    \dot{J}_{\big |_D} = D_2 J = 0
\end{equation} 
\begin{equation}
    \dot{\theta}_{\big |_D} = D_2 \theta = -2|\gamma_{\Delta, \Delta}| \left(\frac{\epsilon \omega_0}{4}\right)^2  \cos(4\theta + 2\Delta t)
\end{equation}
\end{subequations}
For brevity, we kept dependence of the non-Hamiltonian coupling on $\gamma_{\Delta,\Delta}$ implicit. As shown in Fig. \ref{fig:tcg_couplings}, it vanishes on resonance as well. We see that the non-Hamiltonian terms act as phase-dependent modulation of  frequency, pushing the frequency up and down as the phase rotates.

In Fig. \ref{fig:LPTIres} we compare the solution of the problem using our generalized method (or LPT) and using a complete numerical simulation. In this regime of parameters, both the symplectic average of LPT and the convolution average of TCG produce the same effective generator.